\documentclass[aps,a4paper,superscriptaddress,twocolumn]{revtex4}
\usepackage{graphicx,color}
\usepackage{amssymb,amsmath}
\usepackage[hypertexnames=false]{hyperref}
\newcommand{\ILNP}{Laboratory of Nuclear Problems, Joint Institute for Nuclear Research, RU-141980 Dubna, Russia}
\newcommand{\IFZJ}{Institut f\"ur Kernphysik, Forschungszentrum J\"ulich, D-52425 J\"ulich, Germany}
\newcommand{\IGeo}{High Energy Physics Institute, Tbilisi State University, GE-0186 Tbilisi, Georgia}
\newcommand{\IAlm}{Institute of Nuclear Physics, KZ-050032 Almaty, Kazakhstan}
\newcommand{\IAst}{L.N.~Gumilyov Eurasian National University, KZ-010000 Astana, Kazakhstan}
\newcommand{\IGat}{St.\ Petersburg Nuclear Physics Institute, NRC Kurchatov Institute, RU-188350 Gatchina, Russia}
\newcommand{\IMue}{Institut f\"ur Kernphysik, Universit\"at M\"unster, D-48149 M\"unster, Germany}
\newcommand{\IDub}{Dubna State University, RU-141980 Dubna, Russia}
\newcommand{\ISko}{Skobeltsyn Institute of Nuclear Physics, Lomonosov Moscow State University, RU-119991 Moscow, Russia}
\newcommand{\IMSU}{Department of Physics, Moscow State University, RU-119991 Moscow, Russia}
\newcommand{\ILon}{Physics and Astronomy Department, UCL, London WC1E 6BT, United Kingdom}

\begin{document}

\title{Resonance-like coherent production of a pion pair\\in the reaction $\boldsymbol{pd \rightarrow pd\pi\pi}$ in the GeV region}
\author{V.I.~Komarov}\affiliation{\ILNP}
\author{D.~Tsirkov}\email{cyrkov@jinr.ru}\affiliation{\ILNP}
\author{T.~Azaryan}\affiliation{\ILNP}
\author{Z.~Bagdasarian}\affiliation{\IFZJ}\affiliation{\IGeo}
\author{B.~Baimurzinova}\affiliation{\IAlm}\affiliation{\IAst}
\author{S.~Barsov}\affiliation{\IGat}
\author{S.~Dymov}\affiliation{\ILNP}\affiliation{\IFZJ}
\author{R.~Gebel}\affiliation{\IFZJ}
\author{M.~Hartmann}\affiliation{\IFZJ}
\author{A.~Kacharava}\affiliation{\IFZJ}
\author{A.~Khoukaz}\affiliation{\IMue}
\author{A.~Kulikov}\affiliation{\ILNP}
\author{A.~Kunsafina}\affiliation{\ILNP}\affiliation{\IAlm}\affiliation{\IAst}
\author{V.~Kurbatov}\affiliation{\ILNP}
\author{Zh.~Kurmanaliyev}\affiliation{\ILNP}\affiliation{\IAlm}\affiliation{\IAst}
\author{B.~Lorentz}\affiliation{\IFZJ}
\author{G.~Macharashvili}\affiliation{\IGeo}
\author{D.~Mchedlishvili}\affiliation{\IFZJ}\affiliation{\IGeo}
\author{S.~Merzliakov}\affiliation{\IFZJ}
\author{S.~Mikirtytchiants}\affiliation{\IFZJ}\affiliation{\IGat}
\author{M.~Nioradze}\affiliation{\IGeo}
\author{H.~Ohm}\affiliation{\IFZJ}
\author{F.~Rathmann}\affiliation{\IFZJ}
\author{V.~Serdyuk}\affiliation{\IFZJ}
\author{V.~Shmakova}\affiliation{\ILNP}
\author{H.~Str\"oher}\affiliation{\IFZJ}
\author{S.~Trusov}\affiliation{\IFZJ}\affiliation{\ISko}
\author{Yu.~Uzikov}\affiliation{\ILNP}\affiliation{\IDub}\affiliation{\IMSU}
\author{Yu.~Valdau}\affiliation{\IFZJ}\affiliation{\IGat}
\author{C.~Wilkin}\affiliation{\ILon}
\date{\today}
\begin{abstract}
  The reaction $p + d \rightarrow p + d + X$ was studied at 0.8--2.0~GeV proton beam energies with the ANKE magnetic spectrometer at the COSY synchrotron storage ring.
  The proton-deuteron pairs emerging with high momenta, 0.6--1.8~GeV/$c$, were detected at small angles with respect to the proton beam.
  Distribution over the reaction missing mass $M_{x}$ reveals a local enhancement near the threshold of the pion pair production specific for the so-called ABC effect.
  The enhancement has a structure of a narrow bump placed above a smooth continuum.
  The invariant mass of the $d\pi\pi$ system in this enhancement region exhibits a resonance-like peak at ${M}_{d\pi\pi}\approx 2.36$~GeV/$c^2$ with the width $\Gamma\approx0.10$~GeV/$c^2$.
  A possible interpretation of these features is discussed.
\end{abstract}

\maketitle

\section{Introduction}\label{sec:introduction}

Production of a pion pair in $pn$, $pd$ and $dd$ collisions accompanied by emission of a bound light nucleus has been attracting attention since the first observation of a significant local enhancement in the spectrum of the pion pair invariant mass, ${M}_{\pi\pi}$, in the reaction $pd\rightarrow{}^3\mathrm{He}\pi\pi$~\cite{Abashian:1960, Booth:1963}.
The enhancement took place near the threshold of the spectrum, ${M}_{\pi\pi}\sim 300$~MeV/$c^2$, with a surprisingly small width of about $40$~MeV/$c^2$.
This phenomenon got the name of the Abashian-Booth-Crowe (ABC) effect, and its study became a goal of a number of experiments~\cite{Booth:1963, Akimov:1962, Homer:1964, Hall:1969, Banaigs:1973, Bar-Nir:1973, Plouin:1978, Abdivaliev:1980, Holas:1982, Sawada:1997}.
The main established features of the phenomenon were:
\begin{enumerate}\setlength{\itemsep}{0pt}\setlength{\parskip}{0pt}
\item isoscalar nature of the $\pi\pi$ pair,
\item quasi-resonance behavior of the cross section at a fixed scattering angle in relation to the initial energy,
\item strong peaking of the angular distribution in the forward and backward direction,
\item a rather complicated structure of the $\pi\pi$ invariant mass spectra---presence of a wide bump beyond the narrow enhancement,
\item presence of the effect only in reactions accompanied by production of the bound light nucleus, $d$, $^3\mathrm{He}$, $^4\mathrm{He}$.
\end{enumerate}
Some of these features were explained by theoretical models based on dominance of two mechanisms: excitation of two non-interacting $\Delta(1232)$ baryons or excitation of a single Roper baryon $N(1440)$ in the intermediate state of the reactions~\cite{Risser:1973, Bar-Nir:1975, Anjos:1973, Gardestig:1999, AlvarezRuso:1999}.
However, neither of the models could quantitatively describe the whole set of the experimental data.
No convincing explanation was found in particular for the specific feature of the effect, a rather narrow width of the $\pi\pi$ enhancement.

\begin{figure}[htbp]\centering
\includegraphics[width=1\columnwidth]{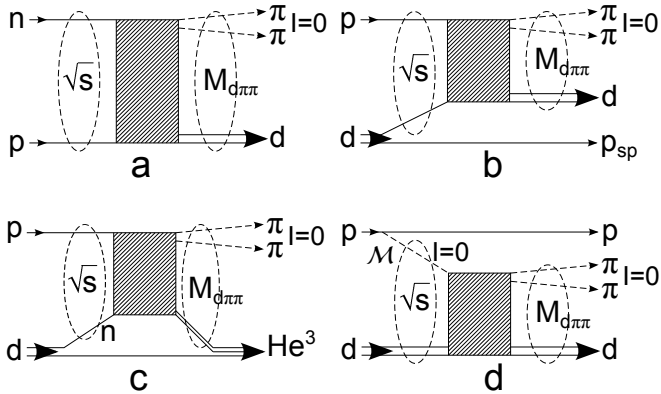}
\caption{Schematic presentation of the two-pion production mechanisms in the reactions $np\rightarrow d\pi\pi$, $pd\rightarrow pd\pi\pi$, and $pd\rightarrow{}^3\mathrm{He}\pi\pi$.
Production of the $d\pi\pi$ system in a free (a) and quasi-free (b) $np$ collision, in formation of the $^3\mathrm{He}\pi\pi$ system (c), and via the meson $\cal{M}$ exchange between the proton and the deuteron (d).
The shaded block depicts an intermediate subprocess leading to the $d\pi\pi$ system formation.
$\sqrt{s}$ denotes the total energy of this subprocess, and $M_{d\pi\pi}$ is the invariant mass of the final $d\pi\pi$ system equal to $\sqrt{s}$.}
\label{fig:scheme.2pi.prod}
\end{figure}

One may suppose that the basis of the ABC phenomenon is the process
\begin{equation}\label{eq:np.to.dpipi}
n+p\rightarrow d+(\pi+\pi)^0
\end{equation}
shown in fig.~\ref{fig:scheme.2pi.prod}a, which enters, as a sub-process, into the amplitude of other reactions with more complicated nuclei (fig.~\ref{fig:scheme.2pi.prod} b, c, d).
However, scantiness of data on reaction~\eqref{eq:np.to.dpipi} due to imperfection of the neutron beams available in 1970s and 1980s limited possible development of theoretical models of the ABC effect.
Furthermore, benefit from use of the proton beams in the reaction
\begin{equation}\label{eq:pd.to.pdpipi}
p+d\rightarrow p+d+(\pi+\pi)^0
\end{equation}
in the regime of the quasi-free interaction of the incoming proton with the neutron inside the deuteron (fig.~\ref{fig:scheme.2pi.prod}b) was constrained by the inclusive character of the experiments~\cite{Homer:1964, Hall:1969}.
A significant increase in interest in the ABC phenomenon in the last decade was caused by the WA\-SA@CELSIUS and WASA@COSY exclusive experiments performed with a high resolution, large acceptance, and robust particle identification setup.
The first study of reaction~\eqref{eq:pd.to.pdpipi} revealed~\cite{Adlarson:2011} a new feature of the ABC effect, namely, the unexpectedly narrow width of a resonance-like peak in the energy dependence of the total cross section of the quasi-free process
\begin{equation}\label{eq:pn.to.dpi0pi0}
p + n\rightarrow d + \pi^{0} +\pi^{0}.
\end{equation}
Within a conventional picture of this process with the $\Delta \Delta$ excitation in the intermediate state one should expect a rather broad width of the peak comparable with twice the width of the free $\Delta(1232)$ isobar, $\Gamma$ = 110~MeV, whereas the observed width is in contrast almost twice as small, about 70~MeV.
Since reaction~\eqref{eq:pn.to.dpi0pi0} proceeds with a relatively high momentum transfer from initial nucleons to nucleons in the final state, it was noted that crucial distances between baryons in the intermediate states are comparable with a characteristic hadron size. It makes the reaction favorable for manifestation of the internal 6$q$ structure of the participating baryon pair~\cite{Adlarson:2014, Bashkanov:2013, Bashkanov:2017}.

Long-term searches for the relevant subject, named ``dibaryon'', used the narrowness of the width of the corresponding resonance, $\Gamma\lesssim100$~MeV/$c^2$, as a basic guideline for identification of the quark structure of the resonance.
For this reason, the resonance with $I(J^P)= 0(3^+)$, the mass of $2.37$~GeV/$c^2$, and $\Gamma\approx70$~MeV observed in the WASA experiments was interpreted~\cite{Bashkanov:2017, Goldman:1989} as clear evidence for the genuine quark-structure dibaryon.
Such interpretation had a theoretical background in the framework of chiral constituent-quark models of the dibaryon (see e.g.~\cite{Goldman:1989, Valcarce:2005, Ping:2009}).
The $0(3^+)$ resonance is also denoted as $D_{03}$ according to the notation $D_{IJ}$ introduced by Dyson and Xuong~\cite{Dyson:1964}, where $I$ is the isospin and $J$ is the angular momentum.

However, the hypothesis of the quark dibaryon nature of the resonance observed in reaction~\eqref{eq:pn.to.dpi0pi0} did not exhaust the ABC problem.
First, interpretation of the effect in terms of the traditional meson-baryon approach is not yet excluded since most of the models used before did not take into account either mutual interaction of the baryons in their intermediate states or coupling of the participating excitation channels.
The studies done last years~\cite{Gal:2013, Gal:2014} resulted in a successful description of some important features of the ABC effect via the meson-baryon models.
Second, the parameters of the $\pi\pi$ yield enhancement observed in~\eqref{eq:pn.to.dpi0pi0} and also in other reactions of the ABC manifestation has no visible connection with the features of the supposed 6$q$ dibaryon.
In particular, the hypothesis does not give a guide for explanation of the detailed behavior of the $\pi\pi$-mass spectrum observed in the SACLAY experimental study~\cite{Banaigs:1973} of the reaction
\begin{equation}\label{eq:dp.to.3Hepipi}
d+p\rightarrow{}^3\mathrm{He}+\pi\pi.
\end{equation}
Descriptions of the perfect data of this experiment in terms of the quark or meson-baryon approach are still unknown.

Clarification of the physical nature of the phenomenon can be advanced by its study in different kinematical conditions.
One of the ways in this direction is variation of the excitation mode of the concerned baryon pair.
In all the experiments so far performed the energy transfer to the pair was realized \textit{via free or quasi-free collision of fast nucleons}.
It is of interest whether the two-baryon resonance and the relevant $\pi\pi$ enhancement could be generated via \textit{coherent interaction of a fast projectile proton with a bound $pn$ pair, the deuteron}.
Here, coherence means, as usual, conservation of the same nucleus in the initial and final states.
In difference to the quasi-free scattering mechanism (fig.~\ref{fig:scheme.2pi.prod}b) and $^3\mathrm{He}\pi\pi$ system production (fig.~\ref{fig:scheme.2pi.prod}c) both nucleons of the deuteron participate in the intermediate subprocess leading to the $d\pi\pi$ system formation (fig.~\ref{fig:scheme.2pi.prod}d).
This subprocess is denoted by the shaded block in fig.~\ref{fig:scheme.2pi.prod}.

\begin{figure*}[htbp]\centering
\includegraphics[width=0.75\textwidth]{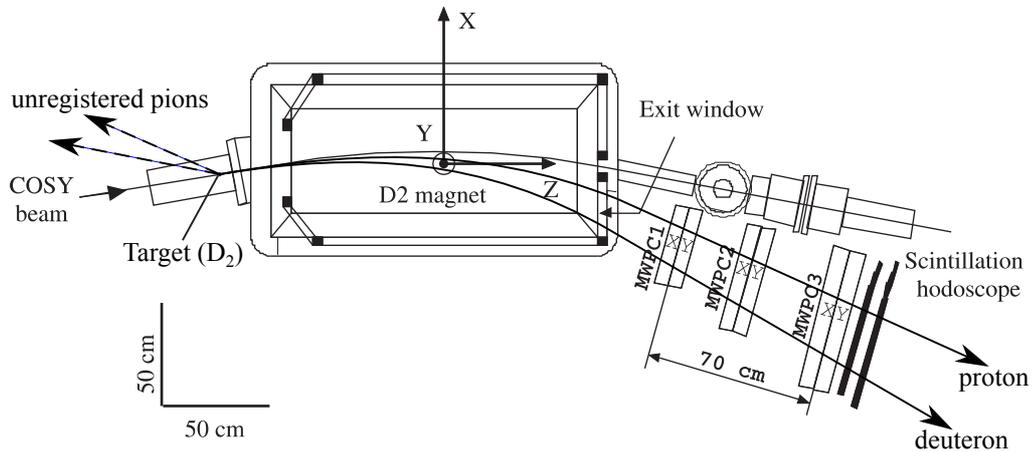}
\caption{The ANKE spectrometer setup (top view), showing the positions of the deuterium cluster jet target and the forward detector (FD).}
\label{fig:ANKE}
\end{figure*}

The deuteron receives the excitation ``as a whole''.
It is clear that such a coherent excitation cannot be achieved by a baryon exchange but occurs via a meson exchange.
Therefore, the kinematical conditions should be favorable for such exchange, that is, the smallest possible transverse momentum transfer and rather limited longitudinal momentum transfer between the initial and final states of the projectile proton.
Such conditions ensure dominance of the simplest mechanism of the process, which is a diagram with a pole in the $t$-channel.
In terms of the kinematical observables it can be achieved in interaction of a proton beam with a deuteron target when both, the proton and the deuteron, are emitted at small angles to the beam with rather high momenta.
A high momentum of the final proton, $p_{p}\geq500$~MeV/$c$, excludes implementation of the quasi-free mechanism of the process, which results in emission of a low-momentum spectator proton (fig.~\ref{fig:scheme.2pi.prod}b).
A high momentum of the deuteron, $p_{d}\geq500$~MeV/$c$, excludes the mechanism of the projectile proton excitation where the deuteron is not exposed to excitation in the intermediate state and therefore gets a rather low momentum.

Earlier studies of the $p+d\rightarrow p+d+\pi+\pi$ reaction performed with a bubble chamber technique (see e.g.~\cite{Brunt:1968, Braun:1976}) did not satisfy the above-stated requirements and were not aimed at studying the ABC effect.
Fortunately, the data obtained at the ANKE@COSY setup in the deuteron break-up study at the beam energies of 0.8--2.0~GeV \cite{Dymov:2010} contained a significant array of events with a final $pd$ pair emitted forward with a high momentum of the both particles.
These events well satisfy the kinematical conditions for the coherent meson production in the reaction
\begin{equation}\label{eq:pd.to.pdX}
p+d\rightarrow p+d+X
\end{equation}
with $X=\pi$, $\pi\pi$, $\eta$ and were processed to study this reaction.
In this paper we present only the two-pion production data of this kind and their analysis.

Therefore, the purpose of this work is an experimental study of the coherent pion pair production in the $pd$ collisions, search for the ABC effect manifestation, and comparison of the results with the known data on this effect.

The experimental setup, technique, and conditions in this work are similar to those used in several experiments performed at ANKE and described in detail before~\cite{Barsov:2001, Chiladze:2002, Dymov:2003, Kurbatov:2008, Dymov:2009, Tsirkov:2010}.
So sect.~\ref{sec:experiment} of the paper gives only a general scheme of the experiment and the features especially significant for the process under consideration.
Data processing, results, and their analysis are given in sect.~\ref{sec:results}.
Section~\ref{sec:discussion} contains discussion of a possible interpretation of the results, and sect.~\ref{sec:summary} summarizes the paper.

\section{Experiment}\label{sec:experiment}

\begin{figure}[htbp]\centering
\includegraphics[width=0.75\columnwidth]{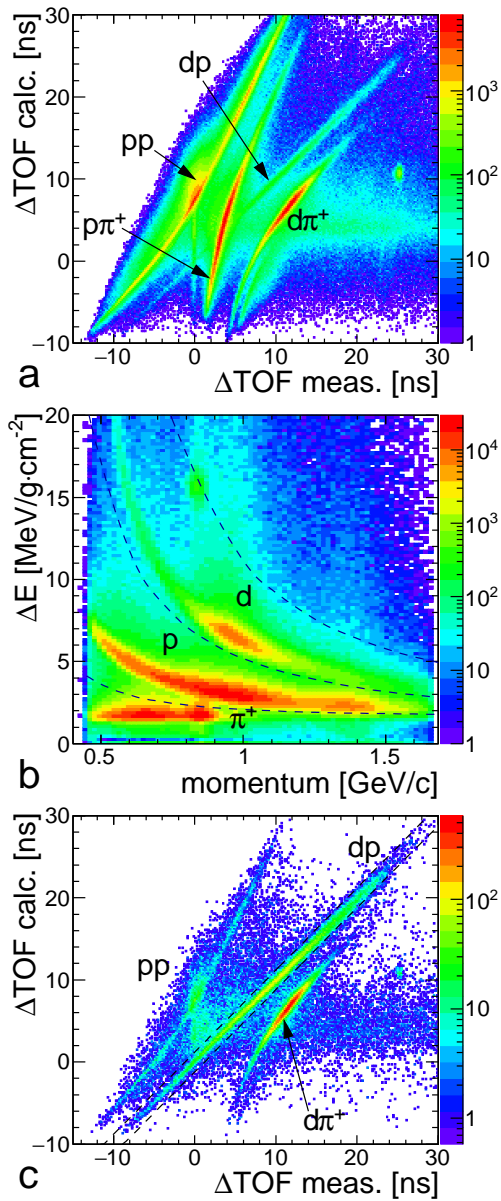}
\caption{Identification of the $pd$ pairs at 1.1~GeV: (a) Distribution of events on the plot of the time-of-flight difference $\Delta \mathrm{TOF}_{\mathrm{meas}}$ measured directly versus the difference $\Delta \mathrm{TOF_{calc}}$ calculated from the momentum measurement; (b) Ionization losses versus momentum, the dashed lines depict the particle separation; (c) $\Delta\mathrm{TOF_{meas}}$ vs.\ $\Delta\mathrm{TOF_{calc}}$ after the use of the ionization loss cuts.
The lines show the final selection of the $pd$ pairs.}
\label{fig:pd.ident}
\end{figure}

The experiment was performed at the proton beam energies $T_{p} =$ 0.8, 1.1, 1.4, and 1.97~GeV with the spectrometer ANKE~\cite{Barsov:2001} installed at the storage ring of the synchrotron COSY.
A scheme of the spectrometer is shown in fig.~\ref{fig:ANKE}.
The beam interacted with the deuterium cluster jet target, and secondary fast protons and deuterons were recorded by multiwire chambers and scintillation counters of the Forward Detector~\cite{Chiladze:2002, Dymov:2003}.
The momenta of the particles were analyzed by the spectrometer magnet.
Recording of at least one particle of the momentum higher than 0.6~GeV/$c$ triggered acquisition of the counter and chamber information.
The angular acceptance of the spectrometer is limited by the forward laboratory angles within $\pm 3.5^{\circ}$ in the vertical and $\pm 12^{\circ}$ in the horizontal plane.
The final particle momentum was measured with an accuracy (RMS) of 0.8\% to 1.2\% for protons and 1\% to 2\% for deuterons.
Precision of the polar angle reconstruction was $0.6^{\circ} (0.8^{\circ})$ for protons (deuterons).
Unambiguous identification of the $pd$ pairs was achieved via measurement of the particle momentum, difference $\Delta\mathrm{TOF}$ of their time of flight from the target to the counters, and ionization energy losses in the counters.
Figure~\ref{fig:pd.ident} illustrates the particle identification at 1.1~GeV.

\begin{figure}[htbp]\centering
\includegraphics[width=1\columnwidth]{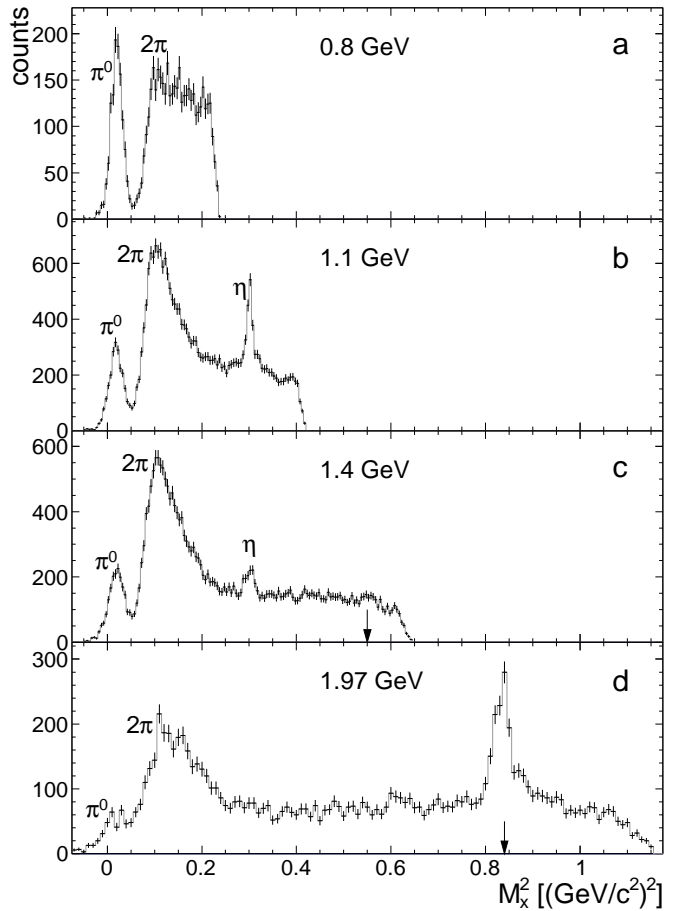}
\caption{
Spectra of the missing mass squared at 0.8~GeV (a), 1.1~GeV (b), 1.4~GeV (c), and 1.97~GeV (d).
Arrows in (c) and (d) mark the position of the $p\pi$ background when the momentum of a pion is falsely accepted as the momentum of a proton.
This background is far from the $0.073$~(GeV/$c^2$)$^2 < M_{X}^2 < 0.17$~(GeV/$c^2$)$^2$ region used further for the two-pion production analysis.
}
\label{fig:MX2}
\end{figure}

Figure~\ref{fig:pd.ident}a shows distribution of the events on the plot $\Delta \mathrm{TOF}_{\mathrm{meas}}$ vs.\ $\Delta \mathrm{TOF}_{\mathrm{calc}}$, where the former is the time difference measured directly and the latter is the difference calculated for particles with the measured momentum, assuming them to be proton and deuteron.
It is seen that the $pd$ pairs are completely separated from the $pp$ and $d\pi$ pairs and mixed with the $p\pi$ pairs only in a small region of the $pd$ and $p\pi$ line intersection.
Strong suppression of the $p\pi$ events and the accidental background was achieved by the appropriate separation of the $pd$ events on the plot of the ionization losses versus momentum (fig.~\ref{fig:pd.ident}b).
The use of the ionization loss cuts provided clean identification of the $pd$ events (fig.~\ref{fig:pd.ident}c).
No admixture of the $p\pi$ events was seen at this energy.
This background arose only at higher energies but was localized in a very limited missing mass $M_X$ region distant from the two-pion region of interest: a small bump near $m_{\pi\pi}^2=0.55$~(GeV/$c^2$)$^2$ at 1.4~GeV and a peak near $m_{\pi\pi}^2=0.84$~(GeV/$c^2$)$^2$ at 1.97~GeV.
It can be seen in fig.~\ref{fig:MX2} c, d.

The luminosity was measured with an accuracy of $\approx$~7\% via simultaneous recording of single protons from small-angle elastic and quasi-elastic scattering off the deuterons.
The data taking, track reconstruction, particle momentum determination and the cross section normalization procedures are described in detail elsewhere~\cite{Dymov:2010, Chiladze:2002, Dymov:2003, Kurbatov:2008}.

Measurement of the 4-momenta of the final proton and deuteron in the reaction $pd\rightarrow pdX$ completely determines kinematics of the reaction.
In particular, the mass $M_X$ and the 3-momentum of the meson system $X$ can be found explicitly.
The $M_{X}^2$ spectra uncorrected for the setup acceptance (fig.~\ref{fig:MX2}~a-d) reveal intense spikes caused by the single $\pi$- and $\eta$-meson production and a broad meson continuum.

\begin{figure}[htbp]\centering
\includegraphics[width=0.8\columnwidth]{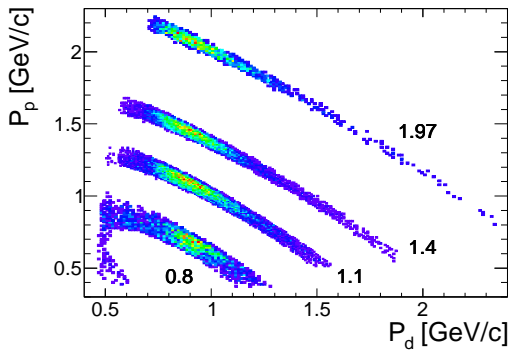}
\caption{The events selected by $M_X^2$ in the $\pi\pi$ peak, $0.073$~(GeV/$c^2$)$^2 < M_X^2 < 0.17$~(GeV/$c^2$)$^2$ at different energies $T_p$ (GeV).
The distribution on the lab momentum plot $P_p$ vs.\ $P_d$.
}
\label{fig:pd.vs.pp}
\end{figure}

\begin{figure}[htbp]\centering
\includegraphics[width=0.9\columnwidth]{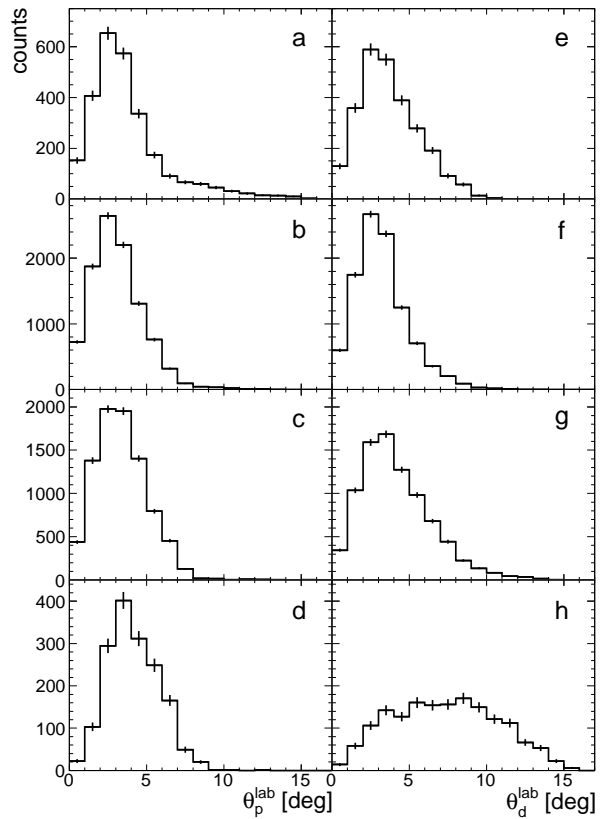}
\caption{The angular distribution of the events over the laboratory angle of the protons $\theta^\mathrm{lab}_p$ (a--d) and the deuterons $\theta^\mathrm{lab}_d$ (e--h).
The upper panels (a, e) correspond to $T_p = 0.8$~GeV; (b, f) to 1.1~GeV; (c, g) to 1.4~GeV; and (d, h) to 1.97~GeV.
}
\label{fig:thetas}
\end{figure}

The meson continuum exhibits a pronounced enhancement at the beam energies of 1.1, 1.4, and 1.97~GeV near the two-pion production threshold in comparison to the smooth multi-pion-production continuum at higher $M_X^2$.
This enhancement at $M_{X}^2 = M_{\pi\pi}^2 \approx 0.1$~(GeV/$c^2$)$^2$ is typical of the ABC effect manifestation.
For further investigation we've selected the region $0.073$~(GeV/$c^2$)$^2 < M_{X}^2 < 0.17$~(GeV/$c^2$)$^2$ that corresponds exactly to pure two-pion production.
Momenta of particles for the events selected in this enhancement region fall into the areas shown in fig.~\ref{fig:pd.vs.pp}.
As is seen, the proton momenta $P_p$ are there higher than 0.6~GeV/$c$ for the proton energies $T_p > 0.8$~GeV, thus excluding the quasi-free regime of the $pn$ interaction.
The deuteron momenta $P_d$ are also higher than this value, which corresponds to significant invariant momentum tra\-nsfers between the initial and the final deuteron: $|t_{dd}| > 0.35$~(GeV/$c$)$^2$.
The angular distribution of the protons is concentrated inside a narrow cone around the initial beam direction (fig.~\ref{fig:thetas}a-d) common to peripheral interactions.
The accompanying deuterons are also emitted at small angles (fig.~\ref{fig:thetas}e-h).

\begin{table*}
\caption{\label{tab:XS.Mdpipi}
The Breit-Wigner parameters of the differential cross section distribution over the invariant mass $M_{d\pi\pi}$ of the deuteron-$\pi\pi$ system: average $\langle M_{d\pi\pi} \rangle$ and width $\Gamma$, along with the process cross section $\Delta\sigma$ within the chosen parameter intervals~\eqref{eq:Mpipi.theta.cuts}.
}
\begin{tabular*}{1\textwidth}{@{\extracolsep{\fill}}ccccccc}
\hline\noalign{\smallskip}
& $T_p$ [GeV] & $\langle M_{d\pi\pi} \rangle\pm\sigma_{st}\pm\sigma_{\mathrm{syst}}$ [GeV/$c^2$] & $\Gamma\pm\sigma_{\mathrm{st}}\pm\sigma_{\mathrm{syst}}$ [GeV/$c^2$] & $\Delta\sigma$ [$\mu$b] & $\chi^2/\mathrm{ndf}$ & \\
\noalign{\smallskip}\hline\noalign{\smallskip}
& $1.1$ & $2.357 \pm 0.002 \pm 0.007$ & $0.115 \pm 0.004 \pm 0.005$ & $0.046 \pm 0.005$ & $29.3/15$ & \\
& $1.4$ & $2.372 \pm 0.002 \pm 0.020$ & $0.092 \pm 0.003 \pm 0.025$ & $0.043 \pm 0.020$ & $19.1/13$ & \\
& averaged & $2.359 \pm 0.001 \pm 0.007 $ & $0.114 \pm 0.001 \pm 0.006$ & $0.046 \pm 0.005$ &  & \\
\noalign{\smallskip}\hline
\end{tabular*}
\end{table*}

\section{Processing, analysis, and results}\label{sec:results}

To find real distributions of the reaction events over kinematical variables, one should correct the observed distributions for the angular-momentum acceptance and detector efficiency of the setup.
The acceptance of the spectrometer can be expressed by a four dimensional matrix in the phase space of the variables $\xi = M_{\pi\pi}$, $M_{d\pi\pi}$, $\theta^\mathrm{cm}_p$, $\theta^{d\pi\pi}_d$, and was determined by the Monte Carlo simulation.
Here $\theta_{p}$ ($\theta_{d}$) is the polar angle of a proton (deuteron) and the superscripts $\mathrm{cm}$ and $d\pi\pi$ stand for the reaction center-of-mass system and the deuteron-two-pion center-of-mass system (CMS), respectively.
The deuteron angle $\theta^{d\pi\pi}_{d}$ was taken in the helicity frame, where the momentum vector of the ejectile proton in the $d\pi\pi$ CMS defined the polar axis.
Azimuthal symmetry of the cross section was assumed, and the setup acceptance was averaged over the azimuthal angles $\phi^{\mathrm{cm}}_p$ and $\phi^{d\pi\pi}_d$.
No acceptance averaging was done over the four variables of the $\xi$ space, thus the acceptance correction was independent on any assumption on the reaction dynamics.
Efficiencies of the detectors were included into the full acceptance factor.
For each of the recorded events, its position in the $\xi$ space was determined and the weight corresponding to the acceptance factor at this point was assigned.
To follow the cross section dependence on $M_{d\pi\pi}$, we integrated the cross section over $M_{\pi\pi}$, $\theta^{\mathrm{cm}}_{p}$ and $\theta^{d\pi\pi}_{d}$ within the fixed limits not varying with $M_{d\pi\pi}$.
These limits were selected by the requirement that the values of the four-dimensional acceptance do not vanish within them in the whole $M_{d \pi\pi}$ range of interest.
For all the beam energies, the integration limits were the same
\begin{equation}\label{eq:Mpipi.theta.cuts}
\begin{array}{r@{~}c@{~}l}
 0.073~(\mathrm{GeV}/c^2)^2 < & M_{\pi\pi}^2 & < 0.17~(\mathrm{GeV}/c^2)^2, \\
 0.982 < & \cos{\theta^\mathrm{cm}_p} & < 1, \\
 -1 < & \cos{\theta^{d\pi\pi}_d} & < -0.98.
\end{array}
\end{equation}
The values of the acceptance varied for the events within these limits from 1.2\% to 69.7\% at 1.1~GeV and from 0.8\% to 68.0\% at 1.4~GeV.

The selected $M_{\pi\pi}^2$ interval encompasses the enhancement region of the pion pairs,
0.27--0.41~GeV/$c^2$, and it means that the further considered data correspond to the reaction
\begin{equation}\label{eq:pd.to.pdpipi0}
p + d \rightarrow p + d + {(\pi\pi)}^{0}.
\end{equation}
The angular intervals correspond to the near-collinear geometry of the proton and $d\pi\pi$ system emission.

\begin{figure}[htbp]\centering
\includegraphics[width=1\columnwidth]{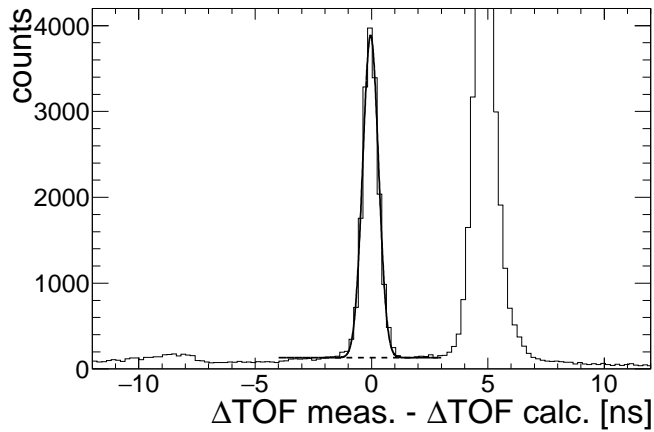}
\caption{
The dependence of counting on the time-delay $\Delta\mathrm{TOF_{meas}} - \Delta\mathrm{TOF_{calc}}$ for the range $8~\mathrm{ns} < \Delta\mathrm{TOF_{meas}} + \Delta\mathrm{TOF_{calc}} < 28~\mathrm{ns}$.
The central peak corresponds to the $dpX$ channel, the right one to the $d\pi^+X$.
The curve represents the fit by the Gaussian plus constant background that gives the height of the Gaussian equal to $(375\pm4)\times10$ and the background intensity $132\pm2$.
}
\label{fig:dTOFdiff}
\end{figure}

The background coming from accidentals and tracking fakes is happily low, $<5\%$.
It can be seen from fig.~\ref{fig:dTOFdiff}, where the dependence of counting on the time-delay $\Delta\mathrm{TOF_{meas}} - \Delta\mathrm{TOF_{calc}}$ is shown for $8~\mathrm{ns} < \Delta\mathrm{TOF_{meas}} + \Delta\mathrm{TOF_{calc}} < 28~\mathrm{ns}$, as most events from the selected $M_{\pi\pi}^2$ interval fall into this range.
Off-target background was measured via switching off the deuteron target jet and was less than 1\%.

\subsection{\texorpdfstring{$\boldsymbol{d\pi\pi}$}{d+pi+pi} and \texorpdfstring{$\boldsymbol{p\pi\pi}$}{p+pi+pi} invariant mass distributions}

Three-momentum transfer from the initial to the final deuteron state larger than 0.6~GeV/$c$ evidently cannot be caused by the quasi-free meson exchange between the projectile and a nucleon inside the deuteron without the deuteron excitation.
Such a process would require too high momentum components of the deuteron wave function.
Another mechanism of the reaction which seems more realistic was discussed in the Introduction and illustrated in fig.~\ref{fig:scheme.2pi.prod}d.
This mechanism is the deuteron coherent excitation via the $t$-channel meson exchange between the projectile proton and the target deuteron.
Absorption of the virtual meson leads to the transfer of significant energy $E_\mathrm{exc} = M^* - M_d$, where $M_d$ is the deuteron mass and $M^*$ is the invariant mass of the excited two-baryon system.
If this excitation is higher than two pion masses, the resulted quasi-stable intermediate state can decay to a pion pair and a deuteron.
Therefore, the invariant mass $M_{d\pi\pi}$ of the produced $d\pi\pi$ system should be equal to the invariant mass of the excited two-baryon system.
The pion pair produced in the decay of the moving excited two-baryon system should be correlated with the other product of the decay, the deuteron.
Feasibility of the $t$-channel meson exchange mechanism in our case is justified by smallness of the transversal 3-momentum transfers and rather low invariant momentum transfers.
Indeed, for the proton laboratory energy from 1--2~GeV and excitation of the deuteron to $E_{\mathrm{exc}} = 0.5$~GeV, the momentum transfer between the initial and the final states of the proton is only 0.45--0.22~GeV/$c$.

\begin{figure}[htbp]\centering
\includegraphics[width=1\columnwidth]{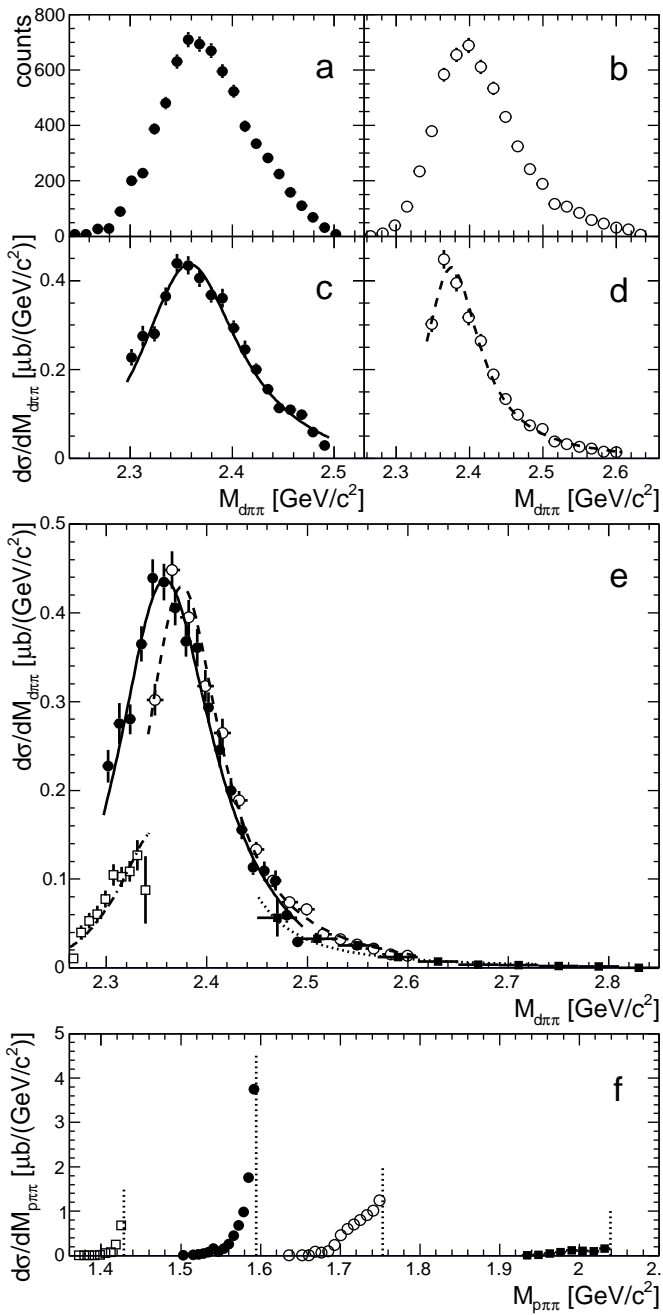}
\caption{
Panels a--b: event counts at $T_p = 1.1$~GeV (a) and 1.4~GeV (b) with the angular and $M_{\pi\pi}$ mass cuts~\eqref{eq:Mpipi.theta.cuts} mentioned in text.
Panels c--d: $M_{d\pi\pi}$ spectra at $T_p = 1.1$~GeV (c) and 1.4~GeV (d) for the regions where the acceptance was calculated precisely, with the same cuts~\eqref{eq:Mpipi.theta.cuts} as in a--b.
The curves show the Breit-Wigner fits.
Panel e: spectra at $T_p =$ 0.8, 1.1, 1.4, and 1.97~GeV are shown together, curves are the Breit-Wigner fits, empty squares and the dash-dotted line correspond to 0.8~GeV, black full circles and the solid line correspond to 1.1~GeV, empty circles and the dashed line correspond to 1.4~GeV, and black full squares and the dotted line correspond to 1.97~GeV.
The Breit-Wigner fits for 0.8 and 1.97~GeV have the mean value fixed at 2.364~GeV/$c^{2}$.
Panel f: spectra of the invariant mass $M_{p\pi\pi}$, the symbols are the same as in (e), vertical dotted lines indicate the kinematical limits.
}
\label{fig:Mdpipi}
\end{figure}

Distributions of the differential cross section over the invariant mass $M_{d\pi\pi}$ are shown in fig.~\ref{fig:Mdpipi}c at 1.1~GeV and in fig.~\ref{fig:Mdpipi}d at 1.4~GeV.
They are corrected for the acceptance in the interval where conditions~\eqref{eq:Mpipi.theta.cuts} are fulfilled and the acceptance can be calculated with a high precision.
The distributions reveal clean peaks inside these intervals.
The raw event counts before correction for the acceptance are shown in fig.~\ref{fig:Mdpipi}a and \ref{fig:Mdpipi}b.
At the edges of the $M_{d\pi\pi}$ range, where only the counts are shown, the acceptance partly vanishes within the limits~\eqref{eq:Mpipi.theta.cuts} and hence could not be calculated in a model-independent way.
Comparing the cross sections and the event counts demonstrates that the calculated acceptances change smoothly and can not produce the peaks artificially.

The acceptance-corrected peaks were fitted with the Breit-Wigner function
\begin{equation}
\frac{d\sigma}{dM_{d\pi\pi}} = \frac{a}{(M_{d\pi\pi} - \langle M_{d\pi\pi}\rangle)^2+{\Gamma^2}/4}
\end{equation}
multiplied by the phase space distribution.
The obtained fit parameters are given in Table~\ref{tab:XS.Mdpipi}.
The experimental FWHM $M_{d\pi\pi}$ resolution is about 12~MeV/$c^2$ and practically does not influence the observed peak widths of $\approx 100$~MeV/$c^2$.
The shown systematic uncertainties of the $\langle M_{d\pi\pi}\rangle$ and $\Gamma$ values are determined mainly by the accuracy of the setup geometry tuning and the choice of the $M_{d\pi\pi}$ region for fitting the data.
The cross section errors also include the uncertainty of the luminosity determination.

While the peak is well observed at the beam energy of 1.1~GeV, its left part is rather poorly determined at 1.4~GeV, resulting in much larger systematic uncertainties of the peak parameters.
The limited acceptance does not allow trailing the whole reso\-nance-like structure at 0.8 and 1.97~GeV: we only get the left part of the peak at 0.8~GeV and the right part at 1.97~GeV.
Nevertheless, these parts do not contradict the assumption that they are tails of the relevant peak with a mean value about 2.36~GeV/$c^2$.
It is clearly seen in fig.~\ref{fig:Mdpipi}e, where the distributions are shown together at all four energies.
The data indicate presence of the same resonance-like system exhibited at the energies used.

In contrast to the resonance behavior of the $d\pi\pi$ system, the $p\pi\pi$ system behaves drastically different in the same $M_{\pi\pi}$ interval.
The events are accumulated at high $M_{p\pi\pi}$ values near the kinematical boundary and do not display any notable resonance structure (fig.~\ref{fig:Mdpipi}f).
It means that the selected kinematical region suppresses excitation of the projectile proton.
If the pion pair is produced by the two-baryon pair excited after the absorption of the $\sigma$-meson, the 3-momentum of the $\pi\pi$ pair should be correlated with the momentum of the final deuteron but not with the scattered projectile proton, which is confirmed experimentally.

The obtained $\langle M_{d\pi\pi}\rangle$ and $\Gamma $ values will be discussed in more detail in sect.~\ref{sec:discussion}.

\subsection{\texorpdfstring{$\boldsymbol{\pi\pi}$}{pi+pi} invariant mass distributions}

\begin{figure}[htbp]\centering
\includegraphics[width=1\columnwidth]{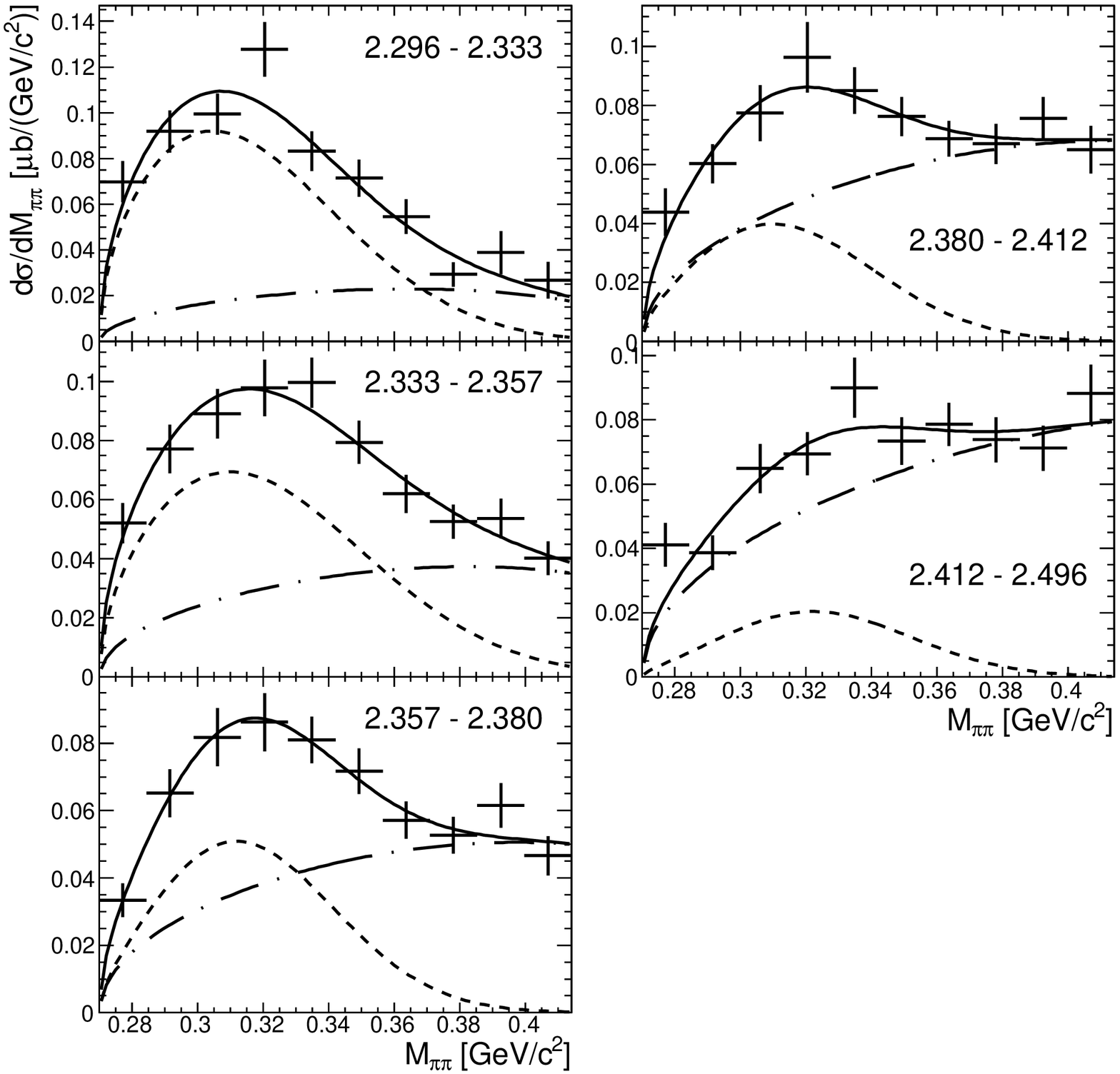}
\caption{
Spectra of the invariant mass $M_{\pi\pi}$ at 1.1~GeV in different $M_{d\pi\pi}$ intervals shown in the panels.
The curves are the Gaussian multiplied by phase space (dotted) and phase space alone (dashed) contributions and their sum (solid), see text.
}
\label{fig:Mpipi.1100.ABC}
\end{figure}

\begin{figure}[htbp]\centering
\includegraphics[width=1\columnwidth]{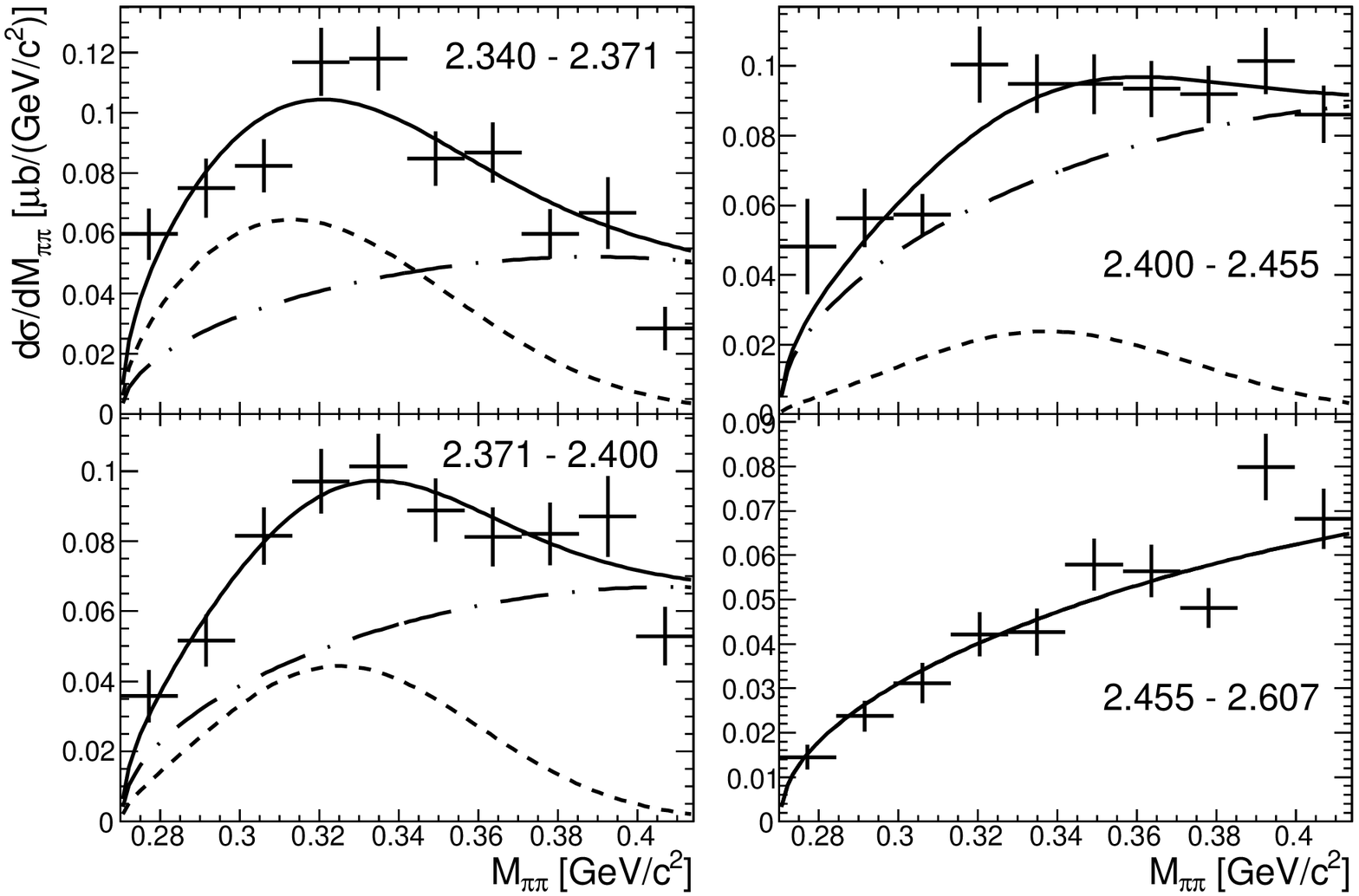}
\caption{
Spectra of the invariant mass $M_{\pi\pi}$ at 1.4~GeV in different $M_{d\pi\pi}$ intervals shown in the panels.
The curves are marked as in fig.~\ref{fig:Mpipi.1100.ABC}.
Note that the Gaussian component vanishes altogether within the 2.455~GeV/$c^2 < M_{d\pi\pi} < 2.607$~GeV/$c^2$ range.
}
\label{fig:Mpipi.1400.ABC}
\end{figure}

\begin{figure}[htbp]\centering
\includegraphics[width=1\columnwidth]{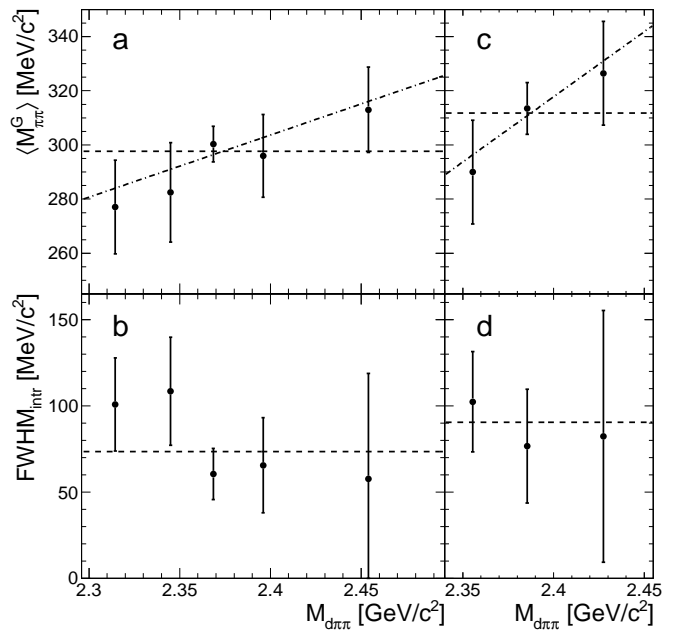}
\caption{
Parameters of the $M_{\pi\pi}$ spectrum at 1.1~GeV (a--b) and 1.4~GeV (c--d) in different $M_{d\pi\pi}$ intervals: (a, c) the mean value $\langle M_{\pi\pi}^{G}\rangle$ and (b, d) the intrinsic width $\mathrm{FWHM_{intr}}$ of the bump, derived from the standard deviation of the Gaussian-shape structure $\sigma(M_{\pi\pi}^{G})$ as described in the text.
The dashed lines mark average values of the parameters.
The dash-dotted line is the linear fit to the $\langle M_{\pi\pi}^{G}\rangle$ points.
}
\label{fig:Mpipi.mean.sigma}
\end{figure}

In fig.~\ref{fig:Mpipi.1100.ABC}--\ref{fig:Mpipi.1400.ABC}, spectra of the pion-pair mass $M_{\pi\pi}$ obtained at $T_p= 1.1$~GeV are shown in different intervals of the $d\pi\pi$ mass.
The intervals scan the $M_{d\pi\pi}$ peak region.
A bump with FWHM $\approx 90$~MeV/$c^2$ is seen above the continuum, smoothly growing with increasing $M_{\pi\pi}$.
The spectra were fitted to the empirical expression
\begin{equation}\label{eq:sigma.G}
d\sigma/dM_{\pi\pi} =(G+1)\Phi,
\end{equation}
where $G$ is the Gaussian
\begin{equation}
G = \left(\sigma(M^{G}_{\pi\pi})\sqrt{2\pi}\right)^{-1} \exp\left(-\frac{\left(M^{G}_{\pi\pi}-\langle M^{G}_{\pi\pi}\rangle\right)^2}{2\sigma^2(M^{G}_{\pi\pi})}\right)
\end{equation}
and $\Phi$ is the phase space.
Form~\eqref{eq:sigma.G} seems to be reasonable since the shape of the bump is essentially determined by the Gaussian shape of the setup resolution with $\mathrm{FWHM}_{\mathrm{res}} = 49$~MeV/$c^2$, which is found from the simulations.
A smoothly varying component taken in a small region near the threshold of the pion pair production is close to the phase space distribution of the pion pair in the $d{\pi\pi}$ system.

Results of such spectrum decomposition are shown by curves in fig.~\ref{fig:Mpipi.1100.ABC}--\ref{fig:Mpipi.1400.ABC}; the $\chi^2/\mathrm{ndf}$ values of the fits vary from $3.1/6$ to $12.5/6$.

Parameters of the Gaussian, $\langle M_{\pi\pi}^G\rangle$ and $\sigma(M_{\pi\pi}^G)$, are presented in fig.~\ref{fig:Mpipi.mean.sigma}.
The mean value of $\langle M_{\pi\pi}^G\rangle$ is $298 \pm 5$~MeV/$c^2$ for 1.1~GeV and $312 \pm 8$~MeV/$c^2$ for 1.4~GeV, while $\langle M_{\pi\pi}^G\rangle$ slightly increases with growing $M_{d\pi\pi}$.
The mean value of $\sigma (M_{\pi\pi}^G)$ is $38 \pm 5$~MeV/$c^2$ for 1.1~GeV and $44 \pm 9$~MeV/$c^2$ for 1.4~GeV, corresponding to the FWHM values $90 \pm 12$ and $103 \pm 21$~MeV/$c^2$.
The intrinsic width of the bump with allowance for the experimental resolution is $\mathrm{FWHM}_{\mathrm{intr}} = 75 \pm 14$~MeV/$c^2$ for 1.1~GeV and $90 \pm 21$~MeV/$c^2$ for 1.4~GeV.
Another way to determine the intrinsic parameters of the bump is to fit the Monte-Carlo simulated spectra to the spectra uncorrected for acceptance and resolution.
In order to calculate the latter, the initial pion pairs were generated in the target according to the $M_{\pi\pi}$ distribution of the $(G+1)\Phi$ shape.
The result of this procedure, $\langle M_{\pi\pi}^G\rangle = 310 \pm 4$~MeV/$c^2$, $\mathrm{FWHM}_{\mathrm{intr}} = 53 \pm 11$~MeV/$c^2$, is slightly different from the previous ones but consistent within the errors.
Therefore, it is reasonable to accept $\mathrm{FWHM}_{\mathrm{intr}} = 65 \pm 8 \pm 8$~MeV/$c^2$, where the first error is statistical and the second one systematic.

\begin{figure}[htbp]\centering
\includegraphics[width=0.8\columnwidth]{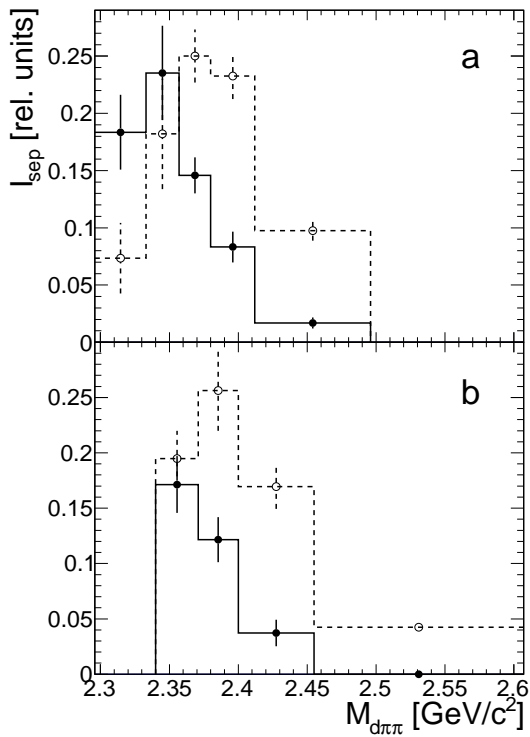}
\caption{
The intensities of the separate components of the $M_{\pi\pi}$ spectra at 1.1~GeV (a) and 1.4~GeV (b) integrated in various intervals of the $d\pi\pi$ invariant mass: full points are the Gaussian component $G\Phi$, and open points are the phase space component $\Phi$.}
\label{fig:Mpipi.components}
\end{figure}

The relative intensity of the Gaussian and the phase space components varies with $M_{d\pi\pi}$.
It is well seen in fig.~\ref{fig:Mpipi.components}, where the distribution of the integrals of these components over $M_{\pi\pi}$ is shown as a function of $M_{d\pi\pi}$: the narrow component prevails at low $M_{d\pi\pi}$ and diminishes with increasing $M_{d\pi\pi}$, while the broad component dominates in the higher $M_{d\pi\pi}$ region.
The $M_{\pi\pi}$ spectra at 1.4~GeV have the similar features.

\section{Discussion}\label{sec:discussion}

As the previous consideration shows, two distinctive peculiarities take place in the double pion production accompanied by the formation of a bound light nucleus:
\begin{enumerate}\setlength{\itemsep}{0pt}\setlength{\parskip}{0pt}
\item resonance-like dependence of the reaction cross section on the $d\pi\pi$ invariant mass,
\item relatively narrow enhancement (peak, hump) in the invariant mass distribution of the final $\pi\pi$ pair.
\end{enumerate}

Several models have been developed for explanation and description of these features, both in traditional me\-son-baryon approaches and in terms of the quark-gluon degrees of freedom.
Here we touch on only those of them which help to understand qualitatively the whole phenomenon.
Discussion of the data from several other ABC-effect experimental studies pursues the same aim.

\subsection{Nature of the \texorpdfstring{$\boldsymbol{d\pi\pi}$}{d+pi+pi} resonance phenomenon}

Let us begin with the resonance-like dependence of the reaction cross section on the $d\pi\pi$ invariant mass.

\begin{figure}[htbp]\centering
\includegraphics[width=0.9\columnwidth]{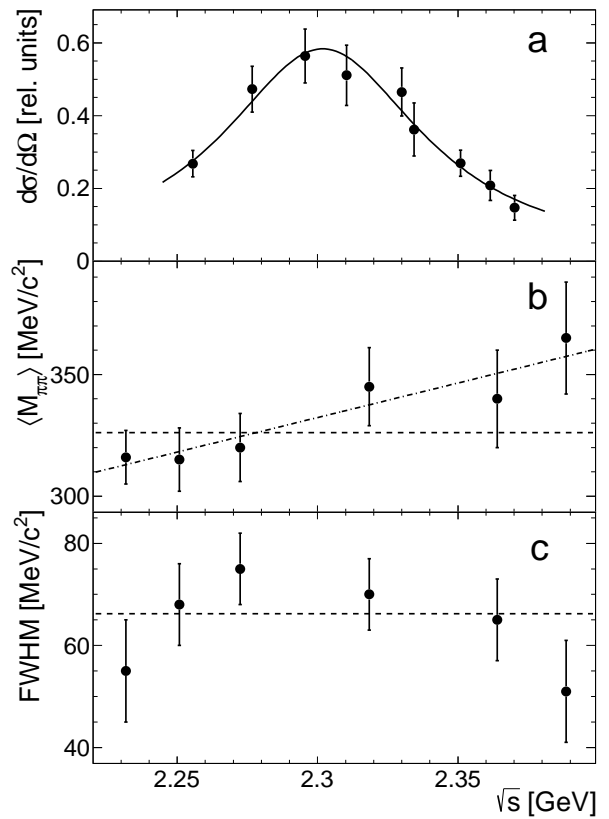}
\caption{
Observables of the reaction $dp\rightarrow{}^3\mathrm{He}\pi\pi$ measured in~\cite{Banaigs:1973} versus the invariant mass $\sqrt{s}$ of the $np\rightarrow d\pi\pi$ subprocess: the differential cross section at the angle of $180^{\circ}$ (a); the mean mass of the $\pi\pi$ pair in the ABC-effect peak (b); the FWHM of the $\pi\pi$ pair ABC peak (c).
The solid line is the Breit-Wigner fit (a), the dashed lines are the fits by a constant (b--c), and the dash-dotted line is the linear fit (c).
}
\label{fig:saclay}
\end{figure}

The first direct observation of the resonance behavior of the cross section was made in the study of reaction~\eqref{eq:dp.to.3Hepipi} in the SACLAY experiment~\cite{Banaigs:1973}.
The differential cross section of the reaction measured at reaction CMS angles near $0^{\circ}$ and $180^{\circ}$ relative to the deuteron beam exhibited a clear resonance dependence on the CMS energy, $W$, with the mean value $W_0 = 3.367$~GeV and the width $\Gamma = 54$~MeV.
The energy $W$ can be related with the full energy $\sqrt{s}$ of the excited intermediate two-baryon system assuming the reaction mechanism shown in fig.~\ref{fig:scheme.2pi.prod}c.
It can be easily done using the kinematical relations relevant for this mechanism.
Making these estimates in the first approximation one can put for the $m_{\pi\pi}$ the mean value of the narrow peak observed in the $m_{\pi\pi}$ distributions~\cite{Banaigs:1973} and assume constant values for the $d\rightarrow p+n$ and $^3\mathrm{He}\rightarrow d+p$ vertices.
Then the kinematics determines the part of the initial deuteron momentum taken away by the neutron in the vertex $d\rightarrow n+p$ and the $\sqrt{s}$ value at a fixed energy $W$.
The obtained dependence of the cross section on the energy $\sqrt{s}$ provides the mean value $\langle\sqrt{s}\rangle = 2.301 \pm 0.003$~GeV and a rather small width $\Gamma = 89 \pm 7$~MeV (fig.~\ref{fig:saclay}a).
The accuracy of these values can be improved by taking into account the form-factors in the $d(pn)$ and $^3\mathrm{He}(dp)$ vertices of the diagram in fig.~\ref{fig:scheme.2pi.prod}c.

Our experimental results can also be compared with those obtained in the recent studies of the ABC effect by the WASA collaboration~\cite{Bashkanov:2009, Adlarson:2011, Adlarson:2014}.
In contrast to our and SACLAY experiments, these studies were done for the exactly determined state of the pion pair, $\pi^0\pi^0$.

The $\langle M_{d\pi\pi}\rangle$ value was estimated to be 2.36~GeV/$c^{2}$ in~\cite{Bashkanov:2009} and 2.37~GeV/$c^{2}$ in~\cite{Kren:2009}.
Later, the mass $M \approx 2.38$~GeV/$c^{2}$ was presented for this resonance~\cite{Adlarson:2011}.
A pole at [$(2380\pm10) - i(40\pm5)$]~MeV was found in the partial wave analysis of the elastic $np$ scattering refined by the measurement of the analyzing power $A_{y}$~\cite{Adlarson:2011}.
It is believed that the pole corresponds to the same $D_{03}$ resonance as was identified in the $pd \to pd\pi^0\pi^0$ reaction study.
The most intriguing feature of the resonance found in the WASA exploration was its rather small width $\Gamma \approx 70$~MeV/$c^{2}$.

The Breit-Wigner mass of the resonance $2.359 \pm 0.007$ GeV/$c^{2}$ observed in our experiment is close to the above-mentioned results with allowance for the systematic errors of the experiments.

However, our value $\Gamma = 114 \pm 6$~MeV/$c^{2}$ is notably higher than the one obtained in~\cite{Bashkanov:2009, Kren:2009, Adlarson:2011}.
It can be assumed that this broadness could be caused by the vertex form factors in the amplitude of the $t$-channel meson exchange (fig.~\ref{fig:scheme.2pi.prod}d).
On the other hand, a possible contribution of the isovector $\pi\pi$ pair in the final state of the reaction~\eqref{eq:pd.to.pdX} can also increase the width of the peak.

Close values of the resonance parameters in the considered experiments and absence of any other resonances in the proximity energy region of these reactions indicates the same nature of the resonances in spite of the different modes of their excitation.
It is reasonable to accept the quantum numbers of the resonances $I(J^{P}) = 0 (3^{+})$ according to the WASA@COSY determination~\cite{Adlarson:2011}.

It is worth noting that the resonance structure with the mass 2.38~GeV and the width about 100~MeV was observed by WASA@COSY in the reaction $pd\to{}^3\mathrm{He}\pi^0\pi^0$ at proton beam energy 1~GeV~\cite{Adlarson:2015}.
Most likely, the origin of the peak is also the $D_{03}$ resonance, while its width is increased due to Fermi-motion of nucleons in the initial deuteron and final $^3\mathrm{He}$ nucleus.

Interpretations of this resonance within constituent-quark models had a significant theoretical support.
A six-quark dibaryon with these quantum numbers and a similar mass was expected in a set of calculations~\cite{Goldman:1989, Valcarce:2005, Ping:2009}.
However, these calculations did not give other important characteristics, such as the width of the resonance, polarization observables, or the cross section of its excitation.
Certainly, the observation~\cite{Adlarson:2014} of the $np$ elastic scattering resonance with the same quantum numbers and energy cannot be evidence in support of the genuine quark nature of the resonance.
A similar relation between elastic and inelastic channels of the resonance were observed much earlier for the strong inelastic isovector $NN$ resonances $^1\!D_2$, $^3\!F_3$, $^3\!P_2$, which had first been discovered in the elastic channels.
Therefore, the traditional meson-baryon approach for interpretation of the resonance associated with the production of the ABC effect remains to be up to date.

Such a study was carried out by Gal and Garcilazo in~\cite{Gal:2013, Gal:2014}.
They employed the $\pi N\Delta$ system in a Faddeev-type three-body calculation, which dynamically generated a pole with its mass and width close to that in the WASA data.
Though the approximation done in the form of a trial $\Delta$ baryon of a zero decay width has lessened the persuasiveness of this result.

More general motivation for narrowing the $D_{03}$ width compared with the free $\Delta(1232)$ width was given recently by Niskanen via a coupled-channel calculation~\cite{Niskanen:2017} with the nucleon, pion, and $\Delta(1232)$ resonance as the only participants in the interaction.
In the last years, development of the chiral constituent-quark model~\cite{Dong:2015, Dong:2016} has resulted in the successful calculation of the total width of the $I(J^{P})$=$0(3^{+})$ resonance with the mass of 2380 MeV.
Moreover, the partial decay widths were reproduced as well.
The calculations revealed a two-component content of the resonance wave function: a $\Delta\Delta$ structure and a dominant hidden-color component, the six-quark exotic state.

In contrast to this, the recent study~\cite{Gal:2017, Gal:2018} done in the meson-baryon concept~\cite{Gal:2013, Gal:2014} describes the total and partial widths of $D_{03}$ decay as a display of the two-component structure of this dibaryon: a compact $\Delta\Delta$ component and a loose $N\Delta$ near-threshold system with $I(J^P)= 1(2^+)$.
This study takes into account a large reduction of the decay width caused by the $\Delta$ motion in the compact $\Delta\Delta$ bound state, the effect missed in the quark-basis calculations~\cite{Dong:2015, Dong:2016}.

Thus, the resonance behavior of the reaction cross section in the region of the ABC effect can be rather successfully reproduced in the framework of the meson-baryon and the quark-model approach.
It means that the term ``dibaryon'' may be used to denote a resonance-like hadr\-on\-ic system with the baryon number two without supposing its fully dense quark-gluon structure.

To summarize, the leading mechanism for the resonance behavior of the reaction cross section associated with the ABC effect is accepted to be the excitation of the $D_{03}$ dibaryon resonance.
Elucidation of the physical nature of the resonance is a particular case of a more general problem to distinguish between a quasi-molecular hadron system and a genuine quark hadron.
This problem was formulated a long time ago in several works (see e.g.\ \cite{Vaughn:1961}).
In~\cite{Weinberg:1963, Weinberg:1965}, Weinberg suggested a way to solve the problem in the case of a bound system, in particular a bound $np$ system, the deuteron.
This way was developed later for unbound states~\cite{Baru:2004, Kamiya:2017}.
It requires several conditions which unfortunately are not met in the dibaryon case under consideration.

Therefore, the existing duality of the dibaryon nature is likely to continue for at least a few coming years.
However, it may serve as a good test bench for exploration of the general ``elementarity-compositeness'' problem.

\subsection{Nature of the ABC narrow \texorpdfstring{$\boldsymbol{\pi\pi}$}{pi+pi}-enhancement}
\label{sec:nature}

Let us now discuss the narrow enhancement in the invariant mass distribution of the final $\pi\pi$ pair.

Decomposition of the $\pi\pi$ invariant mass spectra into a narrow bump and a smooth distribution seen in our data is expressed more distinctly in the SACLAY data~\cite{Banaigs:1973}.
At a relatively low $\sqrt{s}$ corresponding to the left side of the cross section resonance peak a clear narrow peak was seen in the $M_{\pi\pi}$ spectrum with a mean value of $314 \pm 6$~MeV and $\Gamma_{\pi\pi} = 49 \pm 5$~MeV.
An appreciably wider additional component of the spectra arose at $\sqrt{s}$ corresponding to the maximum of the cross section peak, and this component increased in intensity with increasing $\sqrt{s}$ so that on the right side of the resonance the narrow $M_{\pi \pi}$ peak was positioned over a wide distribution of a several times higher intensity.
The mean value and the width of the narrow peak of the ABC effect~\cite{Banaigs:1973} are shown in fig.~\ref{fig:saclay} b, c; the points are taken from Table~2 of \cite{Banaigs:1973}, and the reaction CMS energies are recalculated to the $\sqrt{s}$ values.
They display a general character similar to those observed in our data: a slight rise of the mean value of the peak bump with the energy $\sqrt{s}$ growth and no definite change in the width (fig.~\ref{fig:Mpipi.mean.sigma}).
The same trend in the relative intensity of the narrow and wide components as in~\cite{Banaigs:1973} is seen in our data (fig.~\ref{fig:Mpipi.components}).
Elucidation of the nature of the narrow enhancement and its behavior require a special consideration of the $D_{03}$ decay channels.

From the whole set of the $D_{03}$ decay channels
\begin{subequations}
\begin{align}
D_{03} & \rightarrow N + N \label{eq:D03.a} \\
       & \rightarrow N + N + \pi \label{eq:D03.b} \\
       & \rightarrow N + N + \pi + \pi \label{eq:D03.c} \\
       & \rightarrow d + (\pi +\pi)_{I=0} \label{eq:D03.d} \\
       & \rightarrow d + \sigma \rightarrow d +(\pi + \pi)_{I = 0} \label{eq:D03.e} \\
       & \rightarrow D_{12} + \pi \rightarrow d +(\pi + \pi)_{I = 0} \label{eq:D03.f}
\end{align}
\end{subequations}
channels~\eqref{eq:D03.d}--\eqref{eq:D03.f} are of special interest since they produce exclusively a bound state of a nucleon pair, the deuteron.
$D_{12}$ in~\eqref{eq:D03.f} denotes a well known isovector dibaryon resonance (see e.g.~\cite{Arndt:1987}) with the mass of 2.15~GeV/$c^2$ and quantum number $I(J^{P})$=$1(2^{+})$.
A study of channels~\eqref{eq:D03.e} and~\eqref{eq:D03.f} as an important way of the $d+\pi+\pi$ final state formation was proposed and was done by Platonova and Kukulin~\cite{Platonova:2013}.
Their calculation showed that the $D_{12}+\pi$ channel almost saturated the amplitude of the $D_{03}$ two-body decay mode with the deuteron formation.
Therefore, we limit our consideration to the single mechanism of the reaction
\begin{equation}\label{eq:pd.cascade}
p+d\rightarrow p+ D_{03}\rightarrow p+ D_{12}+\pi_1\rightarrow p+d+\pi_1+\pi_2.
\end{equation}

\begin{figure}[htbp]\centering
\includegraphics[width=0.8\columnwidth]{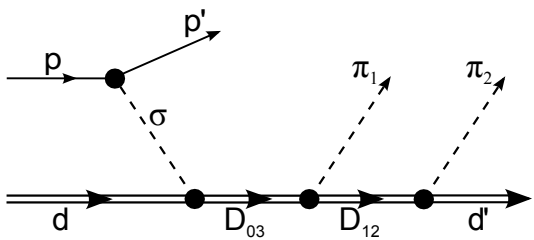}
\caption{The two-step decay mechanism of the reaction $pd\to pd\pi\pi$:
$p+d \to p+D_{03}\to p+D_{12}+\pi_1\to p+d+\pi_1+\pi_2$.
}
\label{fig:2step.decay}
\end{figure}

A two-step decay mechanism of this reaction is schem\-at\-i\-cal\-ly shown in fig.~\ref{fig:2step.decay}.

Channel~\eqref{eq:D03.f} provides a natural explanation for the narrow enhancement in the $m_{\pi\pi}$ distribution.
In this case, the formation of the three-particle final state, $d \pi \pi$, proceeds via the intermediate state of a particle with a well-defined mean value of the mass and a rather narrow full width, the $D_{12}$ dibaryon.
Presence of this particle can lead to a simple kinematical effect of cumulation of the pion pairs in a region with a small relative momentum.
It can be easily seen that if $D_{12}$ decays collinearly along the axis of the $D_{03}$ decay, the pion produced in the decay $D_{03} \rightarrow D_{12} + \pi_1 $ acquires momentum close to that of the second pion produced in the consequent decay $D_{12} \rightarrow d + \pi_2$.
At the mean values $M_{D_{03}} = 2.37$~GeV/$c^{2}$ and $M_{D_{12}} = 2.15$~GeV/$c^{2}$ the $\pi\pi$ invariant mass, that equals 0.271~GeV/$c^{2}$ for $\pi^0\pi^0$ and 0.280~GeV/$c^{2}$ for $\pi^+\pi^-$, becomes nearly a sum of the pion masses.

\begin{figure}[htbp]\centering
\includegraphics[width=0.8\columnwidth]{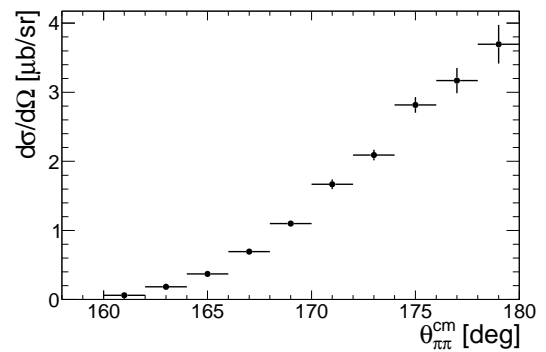}
\caption{
Differential cross section of the $pd \to pd\pi\pi$ reaction within the acceptance limits~\eqref{eq:Mpipi.theta.cuts} against the polar angle $\theta_{\pi\pi}^\mathrm{cm}$ of the pion pair in the reaction CMS, for $T_p = 1.1$~GeV.
}
\label{fig:theta_X}
\end{figure}

The arising kinematical concentration can take place only in the three-particle system $\pi \pi d$, and washes out in the more-participant system $\pi \pi NN$.
This feature explains the well-known fact of the ABC effect manifestation only in the presence of a bound nucleus in the final state.
The existing experimental data strongly intimate the quasi-collinear character of the kinematics associated with the ABC effect: the pion pairs produced in the ABC peaks of the reaction $d + p \rightarrow{}^3\mathrm{He} + (\pi\pi)^{0}$ are distributed in the cone of about $30^{\circ}$--$40^{\circ}$ (FWHM) around $0^{\circ}$ and $180^{\circ}$ relative to the reaction axis (see fig.~20 in~\cite{Banaigs:1973}).
The angular distribution of the deuterons recorded in the WASA experiments also have peaks around the same angles (see fig.~5 in~\cite{Adlarson:2011}).
The kinematics of our experiment is also close to collinear, and the angular distribution of the pion pair is concentrated at high backward angles (fig.~\ref{fig:theta_X}).
Concentration of fast deuterons, $^3\mathrm{He}$, $^4\mathrm{He}$ nuclei near $0^{\circ}$ and $180^{\circ}$ angles is seen in all experiments where the narrow ABC enhancement was observed.

Two factors erode the pion cumulation: receding from the collinear kinematics and dispersion of the resonance masses.
To check to what extent the dispersion of angles and masses can wash out the kinematical cumulation of the pions, we performed a Monte Carlo simulation using the kinematics of the $D_{03}$ and $D_{12}$ decays in channel~\eqref{eq:D03.f}.
The dibaryons were considered to be Breit-Wigner resonances with the parameters $E_R = 2370$~MeV and $\Gamma = 70$~MeV for $D_{03}$, and $E_R = 2120$~MeV and $\Gamma = 120$~MeV for $D_{12}$.
The decay products were emitted into the relevant two-particle phase space.
The resulting $M_{\pi\pi}$ spectrum actually does not depend on the distributions of the $D_{03}$ production and decay, though the concentration of the produced nuclei near $0^{\circ}$ and $180^{\circ}$ suggests that these processes favor collinear kinematics.

\begin{figure}[htbp]\centering
\includegraphics[width=0.8\columnwidth]{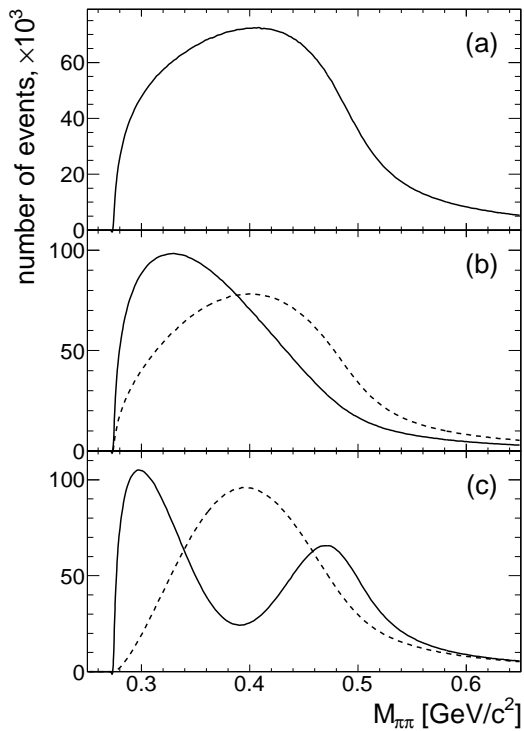}
\caption{
Spectra of the invariant mass $M_{\pi\pi}$ simulated for channel~\eqref{eq:D03.f} with the assumptions described in the text and the angular distributions of the $D_{12}$ decay that are: (a) uniform, (b) solid line for $1 + \cos\theta_d^{D_{12}}$ and dashed line for $1 + \sin\theta_d^{D_{12}}$, (c) solid line for $\cos^2\theta_d^{D_{12}}$ and dashed line for $\sin^2\theta_d^{D_{12}}$.
}
\label{fig:Mpipi.sim}
\end{figure}

Any assumptions on the angular dependence of the $D_{12}$ decay would be unreliable if based on the distributions of final particles without knowing the dynamics of the first two stages of the cascade.
Thus, we have researched the $M^2_{\pi\pi}$ spectra for various terms of a general shape $c_0 + c_1\cos\theta_d^{D_{12}} + c_2\cos^2\theta_d^{D_{12}}$, where $\theta_d^{D_{12}}$ is the polar angle of the deuteron relative to the direction of the $D_{03}$ decay in the $D_{12}$ center-of-mass system.
These simulated spectra are shown in fig.~\ref{fig:Mpipi.sim}.
The uniform $\theta_d^{D_{12}}$ distribution results in a broad and smooth $M_{\pi\pi}$ distribution (fig.~\ref{fig:Mpipi.sim}a).
The distributions in fig.~\ref{fig:Mpipi.sim}b demonstrate influence of the $\theta_d^{D_{12}}$ forward-backward asymmetry on the $M_{\pi\pi}$ distribution.
The distribution shown by a solid line in fig.~\ref{fig:Mpipi.sim}c, obtained with the ``collinear'' angular distribution $\cos^2\theta_d^{D_{12}}$, reveals a pronounced narrow peak resembling ABC effect features.
Its FWHM, $71$~MeV, is almost the same as the typical ABC FWHM of $65\pm11$~MeV.
The ``anti-collinear'' $\sin^2\theta_d^{D_{12}}$ shown by a dashed line, results in a prominent peaking near the center of the spectrum.

Our simulation shows that the narrow $\pi\pi$ enhancement, the ABC effect, may be a consequence of two main peculiarities:
a) the presence of two dibaryon resonances, $D_{03}$ and $D_{12}$, and
b) the decay of $D_{12}$ occurring predominantly in the forward direction.
Confirmation of the suggested explanation of the ABC narrow enhancement undoubtedly requires quantitative calculations, and a thorough theoretical analysis of channel~\eqref{eq:D03.f} is in progress.
The current theory development, e.g., in the ways considered in~\cite{Gal:2013, Platonova:2013, Bashkanov:2017, Clement:2017}, allows expecting such confirmation.
The experimental confirmation can be obtained in the detailed study of the angular dependence of the $M_{\pi\pi}$ spectra: the narrow enhancement should be well seen at the pair emission angles near $0^{\circ}$ and $180^{\circ}$ and should disappear near $90^{\circ}$.

It is natural that the narrow $\pi\pi$ enhancement caused by the kinematical correlation in channel~\eqref{eq:D03.f} is accompanied by a wider excess of the pion pairs from channels~\eqref{eq:D03.d} and \eqref{eq:D03.e}.
An examination in~\cite{Bashkanov:2017} shows in favor of the above a wide set of possible active factors.
A whole structure of the observed $M_{\pi \pi}$ distribution depends evidently on the energy $\sqrt{s}$, the angle of the pair emission, and the momenta of the recorded secondaries from channels~\eqref{eq:D03.d}, \eqref{eq:D03.e}, and \eqref{eq:D03.f}.

In~\cite{Platonova:2013} the experimental $M_{\pi\pi}$ distributions obtained in~\cite{Bashkanov:2009} are well reproduced on the assumption of a significant contribution of channel~\eqref{eq:D03.e}.
For that, the parameters of the $\sigma$ meson must be taken as $M_{\sigma} = 300$~MeV and $\Gamma_{\sigma} = 100$~MeV, strongly different from the currently quoted parameters $(\sqrt{s})^{\sigma}_\mathrm{pole} = ($400--500$) - i($200--300$)$~MeV \cite{Patrignani:2016}.
Such an essential drop in the parameter values was explained as a manifestation of partial chiral symmetry restoration in the conditions of hot and/or dense nuclear matter.
Although the hypothesis of the restoration itself has a serious theoretical support~\cite{Hatsuda:1999,Volkov:1998,Glozman:2000}, its manifestation in the case of the $D_{03}$ resonance decay definitely requires additional confirmations.
Therefore we doubt the use of the sigma-decay channel for explanation of the ABC effect origin and used the channel of the cascade $D_{03} \to D_{12}\pi$ decay.
To conclude, one can consider the wide enhancement in the $M_{\pi \pi}$ distribution as the main contribution of channels~\eqref{eq:D03.d}, \eqref{eq:D03.e}, and partly of \eqref{eq:D03.f}, while the kinematical effect in two-step channel~\eqref{eq:D03.f} can be a source of the narrow enhancement in the $M_{\pi\pi}$ known as the ABC effect discovered in the pioneer works~\cite{Abashian:1960, Booth:1963}.

\section{Summary and conclusion}\label{sec:summary}

\begin{enumerate}\setlength{\itemsep}{0pt}\setlength{\parskip}{0pt}
\item An experimental study of the double pion production at beam energies 0.8--2.0~GeV has been performed in the process $p + d\rightarrow p + d + (\pi \pi)^{0}$ at small scattering angles of the final proton and the deuteron emission.
The final proton momenta are higher than 0.6~GeV/$c$, excluding the quasi-free $np$ scattering mechanism.
The momentum transfers from the initial to the final deuteron are high, 0.4--2~(GeV/$c$)$^2$ in terms of Mandelstam variable $t$, to suppress mechanisms of the projectile proton excitation.
These kinematical conditions are favorable for coherent excitation of the deuteron via the $t$-channel meson exchange between the proton and the deuteron.
\item Significant enlargement of the differential cross section of the reaction at small angles of the proton-deuteron pair emission was observed in the region of the pion pair production close to the threshold.
The distribution of the events over the $d \pi \pi$ invariant mass revealed a clear peak at the mass of 2.36~GeV/$c^{2}$  with the width of about 100~MeV/$c^{2}$.
The parameters of the peak are close to those observed earlier in the study of the double pion production in the quasi-free $NN$ interaction in the $pd \rightarrow pd\pi^0\pi^0$ reaction (WASA at \mbox{CELSIUS}/COSY) and the $dp\rightarrow{}^3\mathrm{He}(\pi\pi)^0$ reaction (SACLAY).
\item The resonance behavior of the coherent production of the pion pair in the $pd\rightarrow pd\pi\pi$ reaction can be considered as a particular case of the $D_{03}$ dibaryon excitation observed in several reactions of the ABC effect manifestation.
The parameters of this dibaryon were determined in the WASA@CELSIUS/COSY studies.
\item The pion pair invariant mass distribution features a two-component character: rather narrow enhancement near 300~MeV/$c^{2}$ with FWHM about 90~MeV/$c^{2}$ over a wide smooth continuum.
This observation is similar to the one obtained in the SACLAY experiment.
\item The pion-pion invariant mass distribution can be explained as manifestation of two leading mechanisms.
One of them is the $D_{03}$ decay channel $D_{03} \rightarrow d + \sigma \rightarrow d + (\pi \pi)_{I=0}$ and the other is $D_{03} \rightarrow D_{12} + \pi \rightarrow d + (\pi \pi )_{I=0}$.
The narrow enhancement called as the ABC effect arises in the second channel due to kinematical correlation between the momenta of the subsequently produced pions.
The arguments presented in favor of this statement are of qualitative nature.
The already developed theoretical models evidently allow the item to be quantitatively examined.
\end{enumerate}

\begin{acknowledgments}
We are grateful to colleagues from the COSY team for providing favorable data-taking conditions.
Valuable discussions with L.~Alvarez-Ruso, V.~Baru, Ch.~Hanhart, A.~Kud\-ryav\-tsev, V.~Kukulin, and J.~Niskanen are acknowledged.
The work was supported in part by the grants of BMBF-JINR, COSY FFE and RFBR (09-02-91332), and by the program of scientific cooperation between JINR  and the Republic of Kazakhstan.
\end{acknowledgments}

\bibliographystyle{apsrev}
\bibliography{pd2pi}

\begin{thebibliography}{55}
\expandafter\ifx\csname natexlab\endcsname\relax\def\natexlab#1{#1}\fi
\expandafter\ifx\csname bibnamefont\endcsname\relax
  \def\bibnamefont#1{#1}\fi
\expandafter\ifx\csname bibfnamefont\endcsname\relax
  \def\bibfnamefont#1{#1}\fi
\expandafter\ifx\csname citenamefont\endcsname\relax
  \def\citenamefont#1{#1}\fi
\expandafter\ifx\csname url\endcsname\relax
  \def\url#1{\texttt{#1}}\fi
\expandafter\ifx\csname urlprefix\endcsname\relax\def\urlprefix{URL }\fi
\providecommand{\bibinfo}[2]{#2}
\providecommand{\eprint}[2][]{\url{#2}}

\bibitem[{\citenamefont{Abashian et~al.}(1960)\citenamefont{Abashian, Booth,
  and Crowe}}]{Abashian:1960}
\bibinfo{author}{\bibfnamefont{A.}~\bibnamefont{Abashian}},
  \bibinfo{author}{\bibfnamefont{N.~E.} \bibnamefont{Booth}}, \bibnamefont{and}
  \bibinfo{author}{\bibfnamefont{K.~M.} \bibnamefont{Crowe}},
  \bibinfo{journal}{Phys. Rev. Lett.} \textbf{\bibinfo{volume}{5}},
  \bibinfo{pages}{258} (\bibinfo{year}{1960}).

\bibitem[{\citenamefont{Booth and Abashian}(1963)}]{Booth:1963}
\bibinfo{author}{\bibfnamefont{N.~E.} \bibnamefont{Booth}} \bibnamefont{and}
  \bibinfo{author}{\bibfnamefont{A.}~\bibnamefont{Abashian}},
  \bibinfo{journal}{Phys. Rev.} \textbf{\bibinfo{volume}{132}},
  \bibinfo{pages}{2314} (\bibinfo{year}{1963}).

\bibitem[{\citenamefont{Akimov et~al.}(1962)\citenamefont{Akimov, Komarov
  et~al.}}]{Akimov:1962}
\bibinfo{author}{\bibfnamefont{{\relax{}Yu}.~K.} \bibnamefont{Akimov}},
  \bibinfo{author}{\bibfnamefont{V.~I.} \bibnamefont{Komarov}},
  \bibnamefont{et~al.}, \bibinfo{journal}{Nucl. Phys.}
  \textbf{\bibinfo{volume}{30}}, \bibinfo{pages}{258} (\bibinfo{year}{1962}).

\bibitem[{\citenamefont{Homer et~al.}(1964)}]{Homer:1964}
\bibinfo{author}{\bibfnamefont{R.~J.} \bibnamefont{Homer}}
  \bibnamefont{et~al.}, \bibinfo{journal}{Phys. Lett.}
  \textbf{\bibinfo{volume}{9}}, \bibinfo{pages}{72} (\bibinfo{year}{1964}).

\bibitem[{\citenamefont{Hall et~al.}(1969)\citenamefont{Hall, Murray, and
  Riddiford}}]{Hall:1969}
\bibinfo{author}{\bibfnamefont{J.~H.} \bibnamefont{Hall}},
  \bibinfo{author}{\bibfnamefont{T.~A.} \bibnamefont{Murray}},
  \bibnamefont{and}
  \bibinfo{author}{\bibfnamefont{L.}~\bibnamefont{Riddiford}},
  \bibinfo{journal}{Nucl. Phys. B} \textbf{\bibinfo{volume}{12}},
  \bibinfo{pages}{573} (\bibinfo{year}{1969}).

\bibitem[{\citenamefont{Banaigs et~al.}(1973)\citenamefont{Banaigs, Berger,
  Goldzahl, Risser, Vu-Hai, Cottereau, and Le~Brun}}]{Banaigs:1973}
\bibinfo{author}{\bibfnamefont{J.}~\bibnamefont{Banaigs}},
  \bibinfo{author}{\bibfnamefont{J.}~\bibnamefont{Berger}},
  \bibinfo{author}{\bibfnamefont{L.}~\bibnamefont{Goldzahl}},
  \bibinfo{author}{\bibfnamefont{T.}~\bibnamefont{Risser}},
  \bibinfo{author}{\bibfnamefont{L.}~\bibnamefont{Vu-Hai}},
  \bibinfo{author}{\bibfnamefont{M.}~\bibnamefont{Cottereau}},
  \bibnamefont{and} \bibinfo{author}{\bibfnamefont{C.}~\bibnamefont{Le~Brun}},
  \bibinfo{journal}{Nucl. Phys. B} \textbf{\bibinfo{volume}{67}},
  \bibinfo{pages}{1} (\bibinfo{year}{1973}).

\bibitem[{\citenamefont{Bar-Nir and et~al}(1973)}]{Bar-Nir:1973}
\bibinfo{author}{\bibfnamefont{I.}~\bibnamefont{Bar-Nir}} \bibnamefont{and}
  \bibinfo{author}{\bibnamefont{et~al}}, \bibinfo{journal}{Nucl. Phys. B}
  \textbf{\bibinfo{volume}{54}}, \bibinfo{pages}{17} (\bibinfo{year}{1973}).

\bibitem[{\citenamefont{Plouin et~al.}(1978)}]{Plouin:1978}
\bibinfo{author}{\bibfnamefont{F.}~\bibnamefont{Plouin}} \bibnamefont{et~al.},
  \bibinfo{journal}{Nucl. Phys. A} \textbf{\bibinfo{volume}{302}},
  \bibinfo{pages}{413} (\bibinfo{year}{1978}).

\bibitem[{\citenamefont{Abdivaliev et~al.}(1980)}]{Abdivaliev:1980}
\bibinfo{author}{\bibfnamefont{A.}~\bibnamefont{Abdivaliev}}
  \bibnamefont{et~al.}, \bibinfo{journal}{Nucl. Phys. B}
  \textbf{\bibinfo{volume}{168}}, \bibinfo{pages}{385} (\bibinfo{year}{1980}).

\bibitem[{\citenamefont{Hollas et~al.}(1982)\citenamefont{Hollas, Newsom,
  Riley, Bonner, and Glass}}]{Holas:1982}
\bibinfo{author}{\bibfnamefont{C.~L.} \bibnamefont{Hollas}},
  \bibinfo{author}{\bibfnamefont{C.~R.} \bibnamefont{Newsom}},
  \bibinfo{author}{\bibfnamefont{P.~J.} \bibnamefont{Riley}},
  \bibinfo{author}{\bibfnamefont{B.~E.} \bibnamefont{Bonner}},
  \bibnamefont{and} \bibinfo{author}{\bibfnamefont{G.}~\bibnamefont{Glass}},
  \bibinfo{journal}{Phys. Rev. C} \textbf{\bibinfo{volume}{25}},
  \bibinfo{pages}{2614} (\bibinfo{year}{1982}).

\bibitem[{\citenamefont{Sawada et~al.}(1997)}]{Sawada:1997}
\bibinfo{author}{\bibfnamefont{S.}~\bibnamefont{Sawada}} \bibnamefont{et~al.},
  \bibinfo{journal}{Nucl. Phys. A} \textbf{\bibinfo{volume}{615}},
  \bibinfo{pages}{277} (\bibinfo{year}{1997}).

\bibitem[{\citenamefont{Risser and Shuster}(1973)}]{Risser:1973}
\bibinfo{author}{\bibfnamefont{T.}~\bibnamefont{Risser}} \bibnamefont{and}
  \bibinfo{author}{\bibfnamefont{M.~D.} \bibnamefont{Shuster}},
  \bibinfo{journal}{Phys. Lett. B} \textbf{\bibinfo{volume}{43}},
  \bibinfo{pages}{68} (\bibinfo{year}{1973}).

\bibitem[{\citenamefont{Bar-Nir et~al.}(1975)\citenamefont{Bar-Nir, Risser, and
  Shuster}}]{Bar-Nir:1975}
\bibinfo{author}{\bibfnamefont{I.}~\bibnamefont{Bar-Nir}},
  \bibinfo{author}{\bibfnamefont{T.}~\bibnamefont{Risser}}, \bibnamefont{and}
  \bibinfo{author}{\bibfnamefont{M.~D.} \bibnamefont{Shuster}},
  \bibinfo{journal}{Nucl. Phys. B} \textbf{\bibinfo{volume}{87}},
  \bibinfo{pages}{109} (\bibinfo{year}{1975}).

\bibitem[{\citenamefont{Anjos et~al.}(1973)\citenamefont{Anjos, Levy, and
  Santoro}}]{Anjos:1973}
\bibinfo{author}{\bibfnamefont{J.~C.} \bibnamefont{Anjos}},
  \bibinfo{author}{\bibfnamefont{D.}~\bibnamefont{Levy}}, \bibnamefont{and}
  \bibinfo{author}{\bibfnamefont{A.}~\bibnamefont{Santoro}},
  \bibinfo{journal}{Nucl. Phys. B} \textbf{\bibinfo{volume}{67}},
  \bibinfo{pages}{37} (\bibinfo{year}{1973}).

\bibitem[{\citenamefont{G\aa{}rdestig et~al.}(1999)\citenamefont{G\aa{}rdestig,
  F\"aldt, and Wilkin}}]{Gardestig:1999}
\bibinfo{author}{\bibfnamefont{A.}~\bibnamefont{G\aa{}rdestig}},
  \bibinfo{author}{\bibfnamefont{G.}~\bibnamefont{F\"aldt}}, \bibnamefont{and}
  \bibinfo{author}{\bibfnamefont{C.}~\bibnamefont{Wilkin}},
  \bibinfo{journal}{Phys. Rev. C} \textbf{\bibinfo{volume}{59}},
  \bibinfo{pages}{2608} (\bibinfo{year}{1999}).

\bibitem[{\citenamefont{Alvarez-Ruso}(1999)}]{AlvarezRuso:1999}
\bibinfo{author}{\bibfnamefont{L.}~\bibnamefont{Alvarez-Ruso}},
  \bibinfo{journal}{Phys. Lett. B} \textbf{\bibinfo{volume}{452}},
  \bibinfo{pages}{207} (\bibinfo{year}{1999}).

\bibitem[{\citenamefont{Adlarson et~al.}(2011)}]{Adlarson:2011}
\bibinfo{author}{\bibfnamefont{P.}~\bibnamefont{Adlarson}} \bibnamefont{et~al.}
  (\bibinfo{collaboration}{WASA-at-COSY Collaboration}),
  \bibinfo{journal}{Phys. Rev. Lett.} \textbf{\bibinfo{volume}{106}},
  \bibinfo{pages}{242302} (\bibinfo{year}{2011}).

\bibitem[{\citenamefont{Adlarson et~al.}(2014)}]{Adlarson:2014}
\bibinfo{author}{\bibfnamefont{P.}~\bibnamefont{Adlarson}} \bibnamefont{et~al.}
  (\bibinfo{collaboration}{WASA-at-COSY Collaboration and SAID Data Analysis
  Center}), \bibinfo{journal}{Phys. Rev. Lett.} \textbf{\bibinfo{volume}{112}},
  \bibinfo{pages}{202301} (\bibinfo{year}{2014}).

\bibitem[{\citenamefont{Bashkanov et~al.}(2013)\citenamefont{Bashkanov,
  Brodsky, and Clement}}]{Bashkanov:2013}
\bibinfo{author}{\bibfnamefont{M.}~\bibnamefont{Bashkanov}},
  \bibinfo{author}{\bibfnamefont{S.~J.} \bibnamefont{Brodsky}},
  \bibnamefont{and} \bibinfo{author}{\bibfnamefont{H.}~\bibnamefont{Clement}},
  \bibinfo{journal}{Phys. Lett. B} \textbf{\bibinfo{volume}{727}},
  \bibinfo{pages}{438} (\bibinfo{year}{2013}).

\bibitem[{\citenamefont{Bashkanov et~al.}(2017)\citenamefont{Bashkanov,
  Clement, and Skorodko}}]{Bashkanov:2017}
\bibinfo{author}{\bibfnamefont{M.}~\bibnamefont{Bashkanov}},
  \bibinfo{author}{\bibfnamefont{H.}~\bibnamefont{Clement}}, \bibnamefont{and}
  \bibinfo{author}{\bibfnamefont{T.}~\bibnamefont{Skorodko}},
  \bibinfo{journal}{Nucl. Phys. A} \textbf{\bibinfo{volume}{958}},
  \bibinfo{pages}{129} (\bibinfo{year}{2017}).

\bibitem[{\citenamefont{Goldman et~al.}(1989)\citenamefont{Goldman, Maltman,
  Stephenson, Schmidt, and Wang}}]{Goldman:1989}
\bibinfo{author}{\bibfnamefont{T.}~\bibnamefont{Goldman}},
  \bibinfo{author}{\bibfnamefont{K.}~\bibnamefont{Maltman}},
  \bibinfo{author}{\bibfnamefont{G.~J.} \bibnamefont{Stephenson}},
  \bibinfo{author}{\bibfnamefont{K.~E.} \bibnamefont{Schmidt}},
  \bibnamefont{and} \bibinfo{author}{\bibfnamefont{F.}~\bibnamefont{Wang}},
  \bibinfo{journal}{Phys. Rev. C} \textbf{\bibinfo{volume}{39}},
  \bibinfo{pages}{1889} (\bibinfo{year}{1989}).

\bibitem[{\citenamefont{Valcarce et~al.}(2005)}]{Valcarce:2005}
\bibinfo{author}{\bibfnamefont{A.}~\bibnamefont{Valcarce}}
  \bibnamefont{et~al.}, \bibinfo{journal}{Rep. Prog. Phys.}
  \textbf{\bibinfo{volume}{68}}, \bibinfo{pages}{965} (\bibinfo{year}{2005}).

\bibitem[{\citenamefont{Ping et~al.}(2009)\citenamefont{Ping, Huang, Pang,
  Wang, and Wong}}]{Ping:2009}
\bibinfo{author}{\bibfnamefont{J.~L.} \bibnamefont{Ping}},
  \bibinfo{author}{\bibfnamefont{H.~X.} \bibnamefont{Huang}},
  \bibinfo{author}{\bibfnamefont{H.~R.} \bibnamefont{Pang}},
  \bibinfo{author}{\bibfnamefont{F.}~\bibnamefont{Wang}}, \bibnamefont{and}
  \bibinfo{author}{\bibfnamefont{C.~W.} \bibnamefont{Wong}},
  \bibinfo{journal}{Phys. Rev. C} \textbf{\bibinfo{volume}{79}},
  \bibinfo{pages}{024001} (\bibinfo{year}{2009}).

\bibitem[{\citenamefont{Dyson and Xuong}(1964)}]{Dyson:1964}
\bibinfo{author}{\bibfnamefont{F.~J.} \bibnamefont{Dyson}} \bibnamefont{and}
  \bibinfo{author}{\bibfnamefont{N.-H.} \bibnamefont{Xuong}},
  \bibinfo{journal}{Phys. Rev. Lett.} \textbf{\bibinfo{volume}{13}},
  \bibinfo{pages}{815} (\bibinfo{year}{1964}).

\bibitem[{\citenamefont{Gal and Garcilazo}(2013)}]{Gal:2013}
\bibinfo{author}{\bibfnamefont{A.}~\bibnamefont{Gal}} \bibnamefont{and}
  \bibinfo{author}{\bibfnamefont{H.}~\bibnamefont{Garcilazo}},
  \bibinfo{journal}{Phys. Rev. Lett.} \textbf{\bibinfo{volume}{111}},
  \bibinfo{pages}{172301} (\bibinfo{year}{2013}).

\bibitem[{\citenamefont{Gal and Garcilazo}(2014)}]{Gal:2014}
\bibinfo{author}{\bibfnamefont{A.}~\bibnamefont{Gal}} \bibnamefont{and}
  \bibinfo{author}{\bibfnamefont{H.}~\bibnamefont{Garcilazo}},
  \bibinfo{journal}{Nucl. Phys. A} \textbf{\bibinfo{volume}{928}},
  \bibinfo{pages}{73} (\bibinfo{year}{2014}).

\bibitem[{\citenamefont{Brunt et~al.}(1968)\citenamefont{Brunt, Clayton, and
  Westwood}}]{Brunt:1968}
\bibinfo{author}{\bibfnamefont{D.~C.} \bibnamefont{Brunt}},
  \bibinfo{author}{\bibfnamefont{M.~J.} \bibnamefont{Clayton}},
  \bibnamefont{and} \bibinfo{author}{\bibfnamefont{B.~A.}
  \bibnamefont{Westwood}}, \bibinfo{journal}{Phys. Lett. B}
  \textbf{\bibinfo{volume}{26}}, \bibinfo{pages}{317} (\bibinfo{year}{1968}).

\bibitem[{\citenamefont{Braun et~al.}(1976)}]{Braun:1976}
\bibinfo{author}{\bibfnamefont{H.}~\bibnamefont{Braun}} \bibnamefont{et~al.},
  \bibinfo{journal}{Nuovo Cimento A} \textbf{\bibinfo{volume}{35}},
  \bibinfo{pages}{45} (\bibinfo{year}{1976}).

\bibitem[{\citenamefont{Dymov et~al.}(2010)}]{Dymov:2010}
\bibinfo{author}{\bibfnamefont{S.}~\bibnamefont{Dymov}} \bibnamefont{et~al.},
  \bibinfo{journal}{Phys. Rev. C} \textbf{\bibinfo{volume}{81}},
  \bibinfo{pages}{044001} (\bibinfo{year}{2010}).

\bibitem[{\citenamefont{Barsov et~al.}(2001)}]{Barsov:2001}
\bibinfo{author}{\bibfnamefont{S.}~\bibnamefont{Barsov}} \bibnamefont{et~al.},
  \bibinfo{journal}{Nucl. Instr. Methods A} \textbf{\bibinfo{volume}{462}},
  \bibinfo{pages}{364} (\bibinfo{year}{2001}).

\bibitem[{\citenamefont{Chiladze et~al.}(2002)}]{Chiladze:2002}
\bibinfo{author}{\bibfnamefont{D.}~\bibnamefont{Chiladze}}
  \bibnamefont{et~al.}, \bibinfo{journal}{Part. Nucl. Lett.}
  \textbf{\bibinfo{volume}{113(4)}}, \bibinfo{pages}{95}
  (\bibinfo{year}{2002}).

\bibitem[{\citenamefont{Dymov et~al.}(2003)}]{Dymov:2003}
\bibinfo{author}{\bibfnamefont{S.}~\bibnamefont{Dymov}} \bibnamefont{et~al.},
  \bibinfo{journal}{Part. Nucl. Lett.} \textbf{\bibinfo{volume}{119(2)}},
  \bibinfo{pages}{40} (\bibinfo{year}{2003}).

\bibitem[{\citenamefont{Kurbatov et~al.}(2008)}]{Kurbatov:2008}
\bibinfo{author}{\bibfnamefont{V.}~\bibnamefont{Kurbatov}}
  \bibnamefont{et~al.}, \bibinfo{journal}{Phys. Lett. B}
  \textbf{\bibinfo{volume}{661}}, \bibinfo{pages}{22} (\bibinfo{year}{2008}).

\bibitem[{\citenamefont{Dymov et~al.}(2009)}]{Dymov:2009}
\bibinfo{author}{\bibfnamefont{S.}~\bibnamefont{Dymov}} \bibnamefont{et~al.},
  \bibinfo{journal}{Phys. Rev. Lett.} \textbf{\bibinfo{volume}{102}},
  \bibinfo{pages}{192301} (\bibinfo{year}{2009}).

\bibitem[{\citenamefont{Tsirkov et~al.}(2010)}]{Tsirkov:2010}
\bibinfo{author}{\bibfnamefont{D.}~\bibnamefont{Tsirkov}} \bibnamefont{et~al.},
  \bibinfo{journal}{J. Phys. G: Nucl. Part. Phys.}
  \textbf{\bibinfo{volume}{37}}, \bibinfo{pages}{105005}
  (\bibinfo{year}{2010}).

\bibitem[{\citenamefont{Bashkanov et~al.}(2009)}]{Bashkanov:2009}
\bibinfo{author}{\bibfnamefont{M.}~\bibnamefont{Bashkanov}}
  \bibnamefont{et~al.}, \bibinfo{journal}{Phys. Rev. Lett.}
  \textbf{\bibinfo{volume}{102}}, \bibinfo{pages}{052301}
  (\bibinfo{year}{2009}).

\bibitem[{\citenamefont{Kren et~al.}(2009)}]{Kren:2009}
\bibinfo{author}{\bibfnamefont{F.}~\bibnamefont{Kren}} \bibnamefont{et~al.},
  \bibinfo{journal}{Int. Journ. of Modern Phys. A}
  \textbf{\bibinfo{volume}{24}}, \bibinfo{pages}{561} (\bibinfo{year}{2009}).

\bibitem[{\citenamefont{Adlarson et~al.}(2015)}]{Adlarson:2015}
\bibinfo{author}{\bibfnamefont{P.}~\bibnamefont{Adlarson}} \bibnamefont{et~al.}
  (\bibinfo{collaboration}{WASA-at-COSY Collaboration}),
  \bibinfo{journal}{Phys. Rev. C} \textbf{\bibinfo{volume}{91}},
  \bibinfo{pages}{015201} (\bibinfo{year}{2015}).

\bibitem[{\citenamefont{Niskanen}(2017)}]{Niskanen:2017}
\bibinfo{author}{\bibfnamefont{J.~A.} \bibnamefont{Niskanen}},
  \bibinfo{journal}{Phys. Rev. C} \textbf{\bibinfo{volume}{95}},
  \bibinfo{pages}{054002} (\bibinfo{year}{2017}).

\bibitem[{\citenamefont{Dong et~al.}(2015)\citenamefont{Dong, Shen, Huang, and
  Zhang}}]{Dong:2015}
\bibinfo{author}{\bibfnamefont{Y.}~\bibnamefont{Dong}},
  \bibinfo{author}{\bibfnamefont{P.}~\bibnamefont{Shen}},
  \bibinfo{author}{\bibfnamefont{F.}~\bibnamefont{Huang}}, \bibnamefont{and}
  \bibinfo{author}{\bibfnamefont{Z.}~\bibnamefont{Zhang}},
  \bibinfo{journal}{Phys. Rev. C} \textbf{\bibinfo{volume}{91}},
  \bibinfo{pages}{064002} (\bibinfo{year}{2015}).

\bibitem[{\citenamefont{Dong et~al.}(2016)\citenamefont{Dong, Huang, Shen, and
  Zhang}}]{Dong:2016}
\bibinfo{author}{\bibfnamefont{Y.}~\bibnamefont{Dong}},
  \bibinfo{author}{\bibfnamefont{F.}~\bibnamefont{Huang}},
  \bibinfo{author}{\bibfnamefont{P.}~\bibnamefont{Shen}}, \bibnamefont{and}
  \bibinfo{author}{\bibfnamefont{Z.}~\bibnamefont{Zhang}},
  \bibinfo{journal}{Phys. Rev. C} \textbf{\bibinfo{volume}{94}},
  \bibinfo{pages}{014003} (\bibinfo{year}{2016}).

\bibitem[{\citenamefont{Gal}(2017)}]{Gal:2017}
\bibinfo{author}{\bibfnamefont{A.}~\bibnamefont{Gal}}, \bibinfo{journal}{Phys.
  Lett. B} \textbf{\bibinfo{volume}{769}}, \bibinfo{pages}{436}
  (\bibinfo{year}{2017}).

\bibitem[{\citenamefont{Gal}(2018)}]{Gal:2018}
\bibinfo{author}{\bibfnamefont{A.}~\bibnamefont{Gal}},
  \bibinfo{journal}{{arXiv}:1803.08788 [nucl-th]}  (\bibinfo{year}{2018}),
  \eprint{1803.08788}.

\bibitem[{\citenamefont{Vaughn et~al.}(1961)\citenamefont{Vaughn, Aaron, and
  Amado}}]{Vaughn:1961}
\bibinfo{author}{\bibfnamefont{M.~T.} \bibnamefont{Vaughn}},
  \bibinfo{author}{\bibfnamefont{R.}~\bibnamefont{Aaron}}, \bibnamefont{and}
  \bibinfo{author}{\bibfnamefont{R.~D.} \bibnamefont{Amado}},
  \bibinfo{journal}{Phys. Rev.} \textbf{\bibinfo{volume}{124}},
  \bibinfo{pages}{1258} (\bibinfo{year}{1961}).

\bibitem[{\citenamefont{Weinberg}(1963)}]{Weinberg:1963}
\bibinfo{author}{\bibfnamefont{S.}~\bibnamefont{Weinberg}},
  \bibinfo{journal}{Phys. Rev.} \textbf{\bibinfo{volume}{130}},
  \bibinfo{pages}{776} (\bibinfo{year}{1963}).

\bibitem[{\citenamefont{Weinberg}(1965)}]{Weinberg:1965}
\bibinfo{author}{\bibfnamefont{S.}~\bibnamefont{Weinberg}},
  \bibinfo{journal}{Phys. Rev.} \textbf{\bibinfo{volume}{137}},
  \bibinfo{pages}{B672} (\bibinfo{year}{1965}).

\bibitem[{\citenamefont{Baru et~al.}(2004)\citenamefont{Baru, Haidenbauer,
  Hanhart, Kalashnikova, and Kudryavtsev}}]{Baru:2004}
\bibinfo{author}{\bibfnamefont{V.}~\bibnamefont{Baru}},
  \bibinfo{author}{\bibfnamefont{J.}~\bibnamefont{Haidenbauer}},
  \bibinfo{author}{\bibfnamefont{C.}~\bibnamefont{Hanhart}},
  \bibinfo{author}{\bibfnamefont{{\relax{}Yu}.}~\bibnamefont{Kalashnikova}},
  \bibnamefont{and}
  \bibinfo{author}{\bibfnamefont{A.}~\bibnamefont{Kudryavtsev}},
  \bibinfo{journal}{Phys. Lett. B} \textbf{\bibinfo{volume}{586}},
  \bibinfo{pages}{53} (\bibinfo{year}{2004}), ISSN \bibinfo{issn}{0370-2693}.

\bibitem[{\citenamefont{Kamiya and Hyobo}(2017)}]{Kamiya:2017}
\bibinfo{author}{\bibfnamefont{Y.}~\bibnamefont{Kamiya}} \bibnamefont{and}
  \bibinfo{author}{\bibfnamefont{T.}~\bibnamefont{Hyobo}},
  \bibinfo{journal}{{arXiv}:1701.0894 [hep-ph]}  (\bibinfo{year}{2017}),
  \eprint{1701.0894}.

\bibitem[{\citenamefont{Arndt et~al.}(1987)\citenamefont{Arndt, Hyslop, and
  Roper}}]{Arndt:1987}
\bibinfo{author}{\bibfnamefont{R.}~\bibnamefont{Arndt}},
  \bibinfo{author}{\bibfnamefont{J.}~\bibnamefont{Hyslop}}, \bibnamefont{and}
  \bibinfo{author}{\bibfnamefont{L.}~\bibnamefont{Roper}},
  \bibinfo{journal}{Phys. Rev. D} \textbf{\bibinfo{volume}{35}},
  \bibinfo{pages}{128} (\bibinfo{year}{1987}).

\bibitem[{\citenamefont{Platonova and Kukulin}(2013)}]{Platonova:2013}
\bibinfo{author}{\bibfnamefont{M.~N.} \bibnamefont{Platonova}}
  \bibnamefont{and} \bibinfo{author}{\bibfnamefont{V.~I.}
  \bibnamefont{Kukulin}}, \bibinfo{journal}{Phys. Rev. C}
  \textbf{\bibinfo{volume}{87}}, \bibinfo{pages}{025202}
  (\bibinfo{year}{2013}).

\bibitem[{\citenamefont{Clement}(2017)}]{Clement:2017}
\bibinfo{author}{\bibfnamefont{H.}~\bibnamefont{Clement}},
  \bibinfo{journal}{Prog. in Part. and Nucl. Physics.}
  \textbf{\bibinfo{volume}{93}}, \bibinfo{pages}{195} (\bibinfo{year}{2017}).

\bibitem[{\citenamefont{Patrignani et~al.}(2016)}]{Patrignani:2016}
\bibinfo{author}{\bibfnamefont{C.}~\bibnamefont{Patrignani}}
  \bibnamefont{et~al.}, \bibinfo{journal}{Chin. Phys. C}
  \textbf{\bibinfo{volume}{40}}, \bibinfo{pages}{100001}
  (\bibinfo{year}{2016}).

\bibitem[{\citenamefont{Hatsuda et~al.}(1999)\citenamefont{Hatsuda, Kunihiro,
  and Shimizu}}]{Hatsuda:1999}
\bibinfo{author}{\bibfnamefont{T.}~\bibnamefont{Hatsuda}},
  \bibinfo{author}{\bibfnamefont{T.}~\bibnamefont{Kunihiro}}, \bibnamefont{and}
  \bibinfo{author}{\bibfnamefont{H.}~\bibnamefont{Shimizu}},
  \bibinfo{journal}{Phys. Rev. Lett.} \textbf{\bibinfo{volume}{82}},
  \bibinfo{pages}{2840} (\bibinfo{year}{1999}).

\bibitem[{\citenamefont{Volkov et~al.}(1998)\citenamefont{Volkov, Kuraev,
  Blaschke, R\"opke, and Schmidt}}]{Volkov:1998}
\bibinfo{author}{\bibfnamefont{M.~K.} \bibnamefont{Volkov}},
  \bibinfo{author}{\bibfnamefont{E.~A.} \bibnamefont{Kuraev}},
  \bibinfo{author}{\bibfnamefont{D.}~\bibnamefont{Blaschke}},
  \bibinfo{author}{\bibfnamefont{G.}~\bibnamefont{R\"opke}}, \bibnamefont{and}
  \bibinfo{author}{\bibfnamefont{S.}~\bibnamefont{Schmidt}},
  \bibinfo{journal}{Phys. Lett. B} \textbf{\bibinfo{volume}{424}},
  \bibinfo{pages}{235} (\bibinfo{year}{1998}).

\bibitem[{\citenamefont{Glozman}(2000)}]{Glozman:2000}
\bibinfo{author}{\bibfnamefont{L.~{\relax{}Ya}.} \bibnamefont{Glozman}},
  \bibinfo{journal}{Phys. Lett. B} \textbf{\bibinfo{volume}{475}},
  \bibinfo{pages}{329} (\bibinfo{year}{2000}).

\end{thebibliography}

\end{document}